\documentclass[fleqn,usenatbib]{mnras}
\usepackage{newtxtext,newtxmath}
\usepackage[T1]{fontenc}
\usepackage{ae,aecompl}

\usepackage{graphicx}	
\usepackage{seqsplit}
\usepackage[all]{hypcap} 
\usepackage{upgreek}
\usepackage{subcaption}
\usepackage{comment}
\usepackage{enumitem}
\usepackage[normalem]{ulem} 

\renewcommand{\eqref}[1]{equation\ (\ref{#1})}
\newcommand{\bba}{$^{\scriptstyle 3\mathrm{D}}$B{\sc arolo}}
\newcommand{\hi}{\ifmmode{\rm HI}\else{H\/{\sc i}}\fi}
\newcommand{\ha}{\ifmmode{\rm H\upalpha}\else{H$\upalpha$}\fi} 
\newcommand{\nii}{$\mathrm{[N\,\textsc{ii}]}$}
\newcommand{\de}{\ifmmode{^\circ}\else{$^\circ$}\fi} 
\newcommand{\vlos}{\ifmmode{V_\mathrm{los}}\else{$V_\mathrm{los}$}\fi}
\newcommand{\vsys}{\ifmmode{V_\mathrm{sys}}\else{$V_\mathrm{sys}$}\fi}
\newcommand{\vrot}{\ifmmode{V_\mathrm{rot}}\else{$V_\mathrm{rot}$}\fi}
\newcommand{\vcirc}{\ifmmode{V_\mathrm{c}}\else{$V_\mathrm{c}$}\fi}
\newcommand{\vflat}{\ifmmode{V_\mathrm{flat}}\else{$V_\mathrm{flat}$}\fi}
\newcommand{\vdisp}{\ifmmode{\sigma_\mathrm{gas}}\else{$\sigma_\mathrm{gas}$}\fi}
\newcommand{\sigmastar}{\ifmmode{\Sigma_\mathrm{\star}}\else{$\Sigma_\mathrm{\star}$}\fi}
\newcommand{\mhalo}{\ifmmode{M_\mathrm{h}}\else{$M_\mathrm{h}$}\fi}
\newcommand{\jstar}{\ifmmode{j_{\star}}\else{$j_{\star}$}\fi}
\newcommand{\fstar}{\ifmmode{f_{\star}}\else{$f_{\star}$}\fi}
\newcommand{\mstar}{\ifmmode{M_{\star}}\else{$M_{\star}$}\fi}
\newcommand{\fj}{\ifmmode{f_\mathrm{j,\star}}\else{f$_\mathrm{j,\star}$}\fi}
\newcommand{\mo}{{\rm M}_\odot}

\newcommand{\kms} {\,{\rm km\,s}^{-1}}

\newcommand{\ang}{\,\text{\rm \AA}}

\newcommand{\galnum}{15}

\usepackage{blindtext}

\title[Dark matter halos of massive spirals]{Dark Matter Halos and Scaling Relations of Extremely Massive Spiral Galaxies from Extended HI Rotation Curves}

\author[Di Teodoro et al.]{Enrico\ M.\ Di Teodoro$^{1,2,3}$\thanks{E-mail: enrico.diteodoro@unifi.it},
Lorenzo Posti$^{4}$, 
S.\ Michael Fall$^{2}$,
Patrick M.\ Ogle$^{2}$,
Thomas Jarrett$^{5}$,
\newauthor
Philip N.\ Appleton$^{6}$,
Michelle E.\ Cluver$^{7}$,
Martha P.\ Haynes$^{8}$,
Ute Lisenfeld$^{9,10}$\\
$^{1}$Department of Physics \& Astronomy, Johns Hopkins University, Baltimore, MD 21218, USA\\
$^{2}$Space Telescope Science Institute, 3700 San Martin Drive, Baltimore, MD 21218, USA\\
$^{3}$Dipartimento di Fisica e Astronomia, Università degli Studi di Firenze, via G.\ Sansone 1, 50019 Sesto Fiorentino, Firenze, Italy\\
$^{4}$Universit\'{e} de Strasbourg, CNRS UMR 7550, Observatoire astronomique de Strasbourg, 11 rue de l’Universit\'{e}, 67000 Strasbourg, France\\
$^{5}$University of Cape Town, Cape Town, South Africa\\
$^{6}$Caltech/IPAC, 1200 E. California Blvd., Pasadena, CA 91125, USA\\
$^{7}$Centre for Astrophysics and Supercomputing, Swinburne University of Technology, Hawthorn, VIC 3122, Australia\\
$^{8}$Cornell Center for Astrophysics and Planetary Science, Cornell University, 122 Sciences Drive, Ithaca, NY 14853, USA \\
$^{9}$Departamento de Física Teórica y del Cosmos, Universidad de Granada, Spain\\
$^{10}$Instituto Carlos I de Física Téorica y Computacional, Facultad de Ciencias, 18071 Granada, Spain
}

\date{Accepted 2022 November 19. Received 2022 November 17; in original form 2022 May 10}

\pubyear{2022}

\begin{document}
\label{firstpage}
\pagerange{\pageref{firstpage}--\pageref{lastpage}}
\maketitle

\begin{abstract}
We present new and archival atomic hydrogen (\hi) observations of \galnum\ of the most massive spiral galaxies in the local Universe ($\mstar>10^{11} \, \mo$).  
From 3D kinematic modeling of the datacubes, we derive extended \hi\ rotation curves, and from these, we estimate masses of the dark matter halos and specific angular momenta of the discs.
We confirm that massive spiral galaxies lie at the upper ends of the Tully-Fisher relation (mass vs velocity, $M \propto V^{4}$) and Fall relation (specific angular momentum vs mass, $j \propto M^{0.6}$), in both stellar and baryonic forms, with no significant deviations from single power laws. 
We study the connections between baryons and dark matter through the stellar (and baryon)-to-halo ratios of mass $f_\mathrm{M} \equiv \mstar/\mhalo$ and specific angular momentum $\fj \equiv \jstar/j_\mathrm{h}$ and $f_\mathrm{j,bar} \equiv j_\mathrm{bar}/j_\mathrm{h}$.  
Combining our sample with others from the literature for less massive disc-dominated galaxies, we find that $f_\mathrm{M}$ rises monotonically with $\mstar$ and $\mhalo$ (instead of the inverted-U shaped $f_\mathrm{M}$ for spheroid-dominated galaxies), while $\fj$ and $f_\mathrm{j,bar}$ are essentially constant near unity over four decades in mass.  
Our results indicate that disc galaxies constitute a self-similar population of objects closely linked to the self-similarity of their dark halos.  
This picture is reminiscent of early analytical models of galaxy formation wherein discs grow by relatively smooth and gradual inflow, isolated from disruptive events such as major mergers and strong AGN feedback, in contrast to the more chaotic growth of spheroids.
\end{abstract}




\section{Introduction}
\label{sec:intro}

Understanding the interplay between baryons and dark matter in galaxy evolution is one of the greatest challenges of modern astrophysics.
In the standard $\Lambda$ cold dark matter ($\Lambda$CDM) cosmological model, galaxies form when gas, trapped within the gravitational potential of dark matter (DM) halos, cools down and collapses toward the centre of the potential well, settling in a disc-like, rotating structure and igniting star formation. 
Subsequently, galaxies grow by means of cosmological accretion and mergers.
While hierarchical assembling of DM halos is reasonably simple and well understood \citep*[e.g.,][]{MovdBW10}, the physics of baryons is complex and involves a number of non-linear physical processes that galaxies experience during their evolution, such as gas dissipation, gas accretion, gravitational instabilities, star formation, feedback from massive stars and active galactic nuclei (AGN), galaxy mergers and tidal interactions.
These processes are expected to alter the properties of a galaxy, such as its mass, rotation velocity and angular momentum.

It is remarkable that galaxies observed in the Universe nonetheless follow tight and approximately featureless power laws between their most basic structural properties \citep[see][for a review]{vanderKruit+2011}.
In particular, the most striking scaling relations for disc galaxies are the \citet{Tully+1977} relation, between the stellar or baryonic\footnote{We follow others in this field and use the term ``baryonic'' as a shorthand for the most readily observable constituents of galaxies, namely stars and atomic gas, thereby excluding molecular and ionized gas, and in particular, the circumgalactic medium.} mass of a galaxy (\mstar\ or $M_\mathrm{bar}$) and its rotation velocity $V$, and the \citet{Fall+1983} relation, between the stellar or baryonic specific angular momentum, $\jstar\equiv J_\star / \mstar$ or $j_\mathrm{bar}\equiv J_\mathrm{bar} / M_\mathrm{bar}$, and the mass, \mstar\ or $M_\mathrm{bar}$.
Both the Tully-Fisher and the Fall relations are power laws, $M \propto V^\beta$ with $\beta\sim3-4$ \citep[e.g.,][]{Verheijen+2001,McGaugh12,Lelli+2016b,Lelli+2019} and $j \propto M^\alpha$ with $\alpha\sim0.5-0.7$ \citep[e.g.,][]{Romanowsky+2012,Fall+2013,Fall+2018,Posti+2018b, ManceraPina+2021,ManceraPina+2021b,DiTeodoro+2021,Hardwick+2022}. 
In a similar way, DM halos in $\Lambda$CDM are fully rescalable and also follow simple, tight, power-law scaling relations with similar slopes, in particular $M_\mathrm{h} \propto V^3_\mathrm{h}$ and $j_\mathrm{h} \propto M^{2/3}_\mathrm{h}$, where $M_\mathrm{h}$, $V_\mathrm{h}$ and $j_\mathrm{h}$ are the mass, circular velocity and specific angular momentum of the dark matter halo, respectively.
This similarity between galaxy and DM scaling relations suggests that a simple connection between dark matter and baryons might still exist in today's disc galaxies, despite all the complexity of galaxy evolution in the cosmological context.

This connection between galaxy and halo properties is not easy to determine from observations \citep{Wechsler+2018}.
A standard way of parametrizing the dependence between baryons and DM is through the ratios $f_{X} \equiv X_\star / X_\mathrm{h}$, i.e. the stellar-to-halo ratio for a given quantity $X$, which can be mass $M$, circular velocity $V$, or specific angular momentum $j$.
In particular, many studies in the last decade have been focused on constraining the variation of $f_\mathrm{M}\equiv \mstar/\mhalo$ or $\fstar = f_\mathrm{M} / f_\mathrm{b}$, where $f_\mathrm{b}\simeq0.156$ is the cosmological baryon fraction \citep{PlanckCollaboration+2020}, as a function of $\mstar$ or $\mhalo$.
This is also referred to as the ``stellar-to-halo mass relation'' (SHMR).
Various techniques have been used so far, including abundance matching \citep[e.g.,][]{Vale+2004,Behroozi+2013,Moster+2013}, weak galaxy-galaxy lensing \citep{Leauthaud+2012,vanUitert+2016}, galaxy clustering \citep{Zu+2015,Tinker+2017}, group catalogs \citep{Zheng+2007,Yang+2008}, and empirical models \citep{Rodriguez-Puebla+2017,Behroozi+2019}.
The general consensus is that $f_\mathrm{M}$ (or \fstar) increases with mass, reaches a peak around $\mstar\simeq5\times10^{10} \, \mo$ ($\mhalo\simeq10^{12} \, \mo$, near the mass of the Milky Way), and then decreases at larger masses.
The standard interpretation of this form of the SHMR is that 
the rising and falling parts reflect mainly stellar and AGN feedback, respectively.
A puzzling aspect of such a non-linear\footnote{For brevity, we often use the terms ``linear'' and ``non-linear'' to refer to the power-law and non-power-law forms of relations in log-log space.} relation is that a bend in the SHMR should induce a bend in the disc scaling relations at about the same mass \citep[e.g.,][]{Navarro+00,Ferrero+2017,Lapi+2018,Posti+2018a}. 
Such a bend, however, is not observed \citep{DiTeodoro+2021}.

A simple and elegant solution to this puzzle was recently put forward by \citet{Posti+2019b,Posti+2019} and \citet{PostiFall21}: that early-type galaxies (spheroids) and late-type galaxies (discs) in reality follow different SHMRs.
In particular, \citet{Posti+2019} estimated DM halo masses from extended atomic hydrogen (\hi) rotation curves in disc galaxies and found that the SHMR rises monotonically for all masses, at least up to $\mstar\sim10^{11} \, \mo$. 
Conversely, \citet{PostiFall21} derived the SHMR for early-type galaxies from position-velocity distributions of their globular clusters and found a SHMR that declines above $\mstar\sim5\times10^{10} \, \mo$, in agreement with the standard SHMR derived from abundance matching for the general population of galaxies, which is dominated by early-types at the highest masses \citep[e.g.,][]{Kelvin+2014}.
The linearity of the SHMR for discs suggests that, unlike in spheroids, AGN feedback is not efficient enough to turn over the SHMR relation (``failed feedback" problem).
This likely happens because the super-massive black holes that power AGN feedback are associated more with spheroids than with discs \citep{Kormendy+1995}.
Differences in the SHMR of early-type and late-type galaxies have also been found in recent hydrodynamical cosmological simulations \citep{Grand+2019,Marasco+2020,Correa+2020,Rodriguez-Gomez+2022}.
These findings indicate that massive discs and spheroids are assembled in different ways, with the former being built by smooth inflows regulated by relatively weak stellar feedback, and the latter being assembled by a combination of merging and strong AGN feedback.

For disc-dominated galaxies, the other important property, next to mass, is specific angular momentum. This must ultimately derive from the tidal torques exerted by density perturbations on a wide range of scales in the linear and trans-linear phases of galaxy formation \citep{Peebles+1969}. 
As first emphasized by \citet{Fall+1980}, the torques, being gravitational, would act nearly equally on both the gas and dark matter in protogalaxies, thus endowing them initially with approximately the same specific angular momentum; and for discs especially, this equality would largely be preserved during the subsequent collapse and later evolution of galaxies.  
The assumption that galactic discs have the same specific angular momentum as their dark halos underlies practically all analytical and semi-analytical models of galaxy formation developed over the past forty years.  
There are, however, several processes that can alter this relationship, including transfer of angular momentum from galaxies to their dark halos, major mergers and tidal interactions with neighboring galaxies, and removal of gas by stellar and AGN feedback — thought to be dominant for spheroids but subdominant for discs (reviewed by \citealt{Romanowsky+2012}; see also \citealt{DeFelippis+2017}).
Thus, it is important to make robust empirical estimates of the ``retained fractions'' of specific angular momentum, $\fj\equiv \jstar/j_\mathrm{h}$ and $f_\mathrm{j,bar}\equiv j_\mathrm{bar}/j_\mathrm{h}$ for the stars and baryons in galaxies over wide ranges in $\mstar$ and $\mhalo$. 
Recent estimates are $\fj \simeq 0.7-0.8$ for disc-dominated galaxies \citep{Fall+2013,Fall+2018,Huang+2017,Posti+2019,DiTeodoro+2021} and  $\fj \simeq 0.1$ for spheroid-dominated galaxies \citep{Romanowsky+2012,Fall+2018}, in general agreement with recent cosmological hydrodynamical simulations \citep[][]{Rodriguez-Gomez+2022}.  

In this paper, we aim to study the connection between very massive discs and their DM halos, extending the stellar mass range well above $10^{11} \, \mo$, and to test their growth scenario.
We use a combination of new and archival \hi\ data to derive extended rotation curves of a sample of \galnum\ galaxies that are among the most massive discs in the local Universe.
These rotation curves are then decomposed into the contribution of the different mass components (cold gas, stars and dark matter) and robust estimates of DM halo masses and specific angular momenta are obtained by imposing physically-motivated priors on the concentration and spin parameters of $\Lambda$CDM halos.
We stress that \hi\ rotation curves, which extend well beyond the stellar discs of galaxies, are crucial for estimating DM halo masses \citep[e.g.,][]{Bosma+1978,vanAlbada+1985,Kent+1988}.
Finally, we build and study the Tully-Fisher and Fall relations, in both their stellar and baryonic flavors, the SHMRs and the relation between stellar and halo specific angular momenta for disc galaxies with $\mstar\gtrsim10^{11} \, \mo$. 

The remainder of this paper is structured as follows. 
Section~\ref{sec:data} introduces our galaxy sample and the kinematical emission-line data used in this work, including our new \hi\ observations. 
Section~\ref{sec:methods} describes our 3D kinematical modelling of the data and our rotation curve decomposition procedure to estimate the mass of dark matter halos in our galaxies.
In Section~\ref{sec:results}, we present and discuss our results on scaling relations and dark matter halos of extremely massive spiral galaxies. 
The physical implications of our findings are discussed in Section~\ref{sec:disc}.
We summarize and conclude in Section~\ref{sec:conclusions}.
Throughout the paper, we use a fixed critical overdensity parameter $\Delta=200$ for DM halos and we assume a flat $\Lambda$CDM cosmology with $\Omega_\mathrm{m} = 0.314$, 
$\Omega_\mathrm{\Lambda} = 0.686$ and $H_0 = 67.4$ km s$^{-1}$ Mpc$^{-1}$ \citep{PlanckCollaboration+2020}.
In this cosmology, 1 arcsec corresponds to 212 pc and the lookback time is 143 Myr at $z \simeq 0.01$.

\section{Data}
\label{sec:data}

\subsection{Galaxy sample and atomic hydrogen data}
\label{sec:HIdata}

This work uses a mixture of new and archival \hi\ interferometric data. Of the \galnum\ galaxies that make up our final sample, 7 come from new observations (out of 8 galaxies actually observed) and 8 from previous published studies. 

For this project, we first selected a sample of 8 new massive spiral galaxies to be observed with 
the Karl Jansky Very Large Array (VLA).
These galaxies were chosen based on the their stellar masses, their rotational velocities and their inclination angles.
Galaxies were required to have stellar mass $\mstar>10^{11}$, based on the Sloan Digital Sky Survey \citep[SDSS,][]{Eisenstein+2011} catalog, and \hi\ rotation speeds $\vrot>300 \, \kms$, based on their inclination-corrected \hi\ line profile width from the Arecibo ALFALFA survey \citep{Haynes+2018}.
We note that ALFALFA galaxies typically have higher gas fractions, are more isolated and have higher gas spin parameters than the general galaxy population (e.g.\ SDSS).
Finally, we required an inclination angle $i>40^\circ$ to minimize uncertainties on the kinematical modelling due to projection effects.

Aperture synthesis \hi\ data for these galaxies were collected with the VLA in separate runs from July to September 2021 (project ID: 21A-191, P.I.: Ogle).
The telescope was in configuration C, with a minimum baseline of 35 m and a maximum baseline of 3.4 km.
A simple spectral setup for the L-band receiver was used, with a 16 MHz spectral window centred on the redshifted \hi\ line and 512 channels of width 31.25 kHZ, corresponding to about 6.5 $\kms$ at the \hi\ frequencies. 
This configuration guarantees sufficient bandwidth to include the broad integrated \hi\ emission for these fast-rotating galaxies as well as enough baseline to subtract continuum emission. 
We used a standard observing sequence, in which a galaxy was targeted in $20-30$ minute intervals, separated by $3-5$ minute phase calibrator pointings. 
The flux/bandpass calibrator was observed at the start and at the end of each run. 
The total on-source integration time for each galaxy is approximately 9 hours.

Calibration and imaging of the interferometric radio data was performed using the Common Astronomy Software Applications \citep[CASA,][]{McMullin+2007}. 
In particular, we used stable version 6.1.
Raw data were accurately cleaned using both automated algorithms (TFCrop or RFlag) and manual flagging to remove undesired radio-frequency interference (RFI) and other bad data. 
For each observation run, we used standard CASA routines to perform flux, bandpass and complex gain calibrations.
Calibration solutions and calibrated visibilities were visually inspected to ensure high-quality data for imaging.
To increase the signal-to-noise ratio (SNR), calibrated sets were binned by either two or three channels, depending on the galaxy, achieving a channel width of 62.5 kHz or 93.75 kHz ($\sim13 \, \kms$ and $\sim20 \, \kms$, respectively).

Continuum emission was removed from calibrated data in the $uv$-plane by fitting a first order polynomial to \hi\ emission-free channels and subtracting it from visibility spectra over the entire spectral range.
Continuum-subtracted spectra were imaged on a grid with pixel spacing of $4''$ using a natural weighting scheme to maximize the sensitivity of data.
Finally, the emission in datacubes was cleaned of sidelobe contamination \citep{Clark+1980} down to a $\mathrm{SNR}=3$ and corrected for the attenuation of the primary beam.
Our final datacubes reach a typical root mean square (rms) sensitivity of $\sigma_\mathrm{rms} \simeq 0.2-0.4$ mJy beam$^{-1}$ per channel and have typical Full Width at Half Maximum (FWHM) beam sizes of $17''-21''$.
Of the 8 new galaxies observed with the VLA, we had to discard UGC08902, which turned out to be a very strongly interacting system, thus unsuitable for our dynamical analysis. 

Beside these new VLA data, we additionally re-analysed archival \hi\ data of 8 massive, fast-rotating, disc galaxies. 
In particular, we used VLA data from \citet{Spekkens+2006} for 3 galaxies (NGC1324, NGC2862 and UGC2849),  and Westerbork Synthesis Radio Telescope (WSRT) data from the Westerbork survey of neutral Hydrogen in Irregular and SPiral galaxies \citep[WHISP,][]{vanderHulst+2001} for 4 systems (NGC338, NGC1167, NGC2599 and NGC5533). 
Finally, we also analysed WSRT data for UGC02885 from \citet{Hunter+2013}, also known as ``Rubin's galaxy".
Three of these galaxies (NGC1167, NGC5533 and UGC02885) are also included in the Spitzer Photometry \& Accurate Rotation Curves catalog \citep[SPARC,][]{Lelli+2016b}, but we decided to re-analyse these data with our latest modelling techniques.
Archival datasets have beam FWHMs of $14''-45''$ and channel widths of $4-13$ $\kms$, while the rms sensitivity varies across the sample in the range $\sigma_\mathrm{rms}=0.3-1.1$ mJy beam$^{-1}$ in a single channel.

Galaxies in our final sample have stellar masses that exceed $10^{11} \, \mo$, distances $D<250$ Mpc ($z\lesssim0.05$) and morphologies spanning from early-type discs (e.g. S0, Sa) to late-type discs (e.g., Sb, Sc).
These galaxies are lower-redshift analogs of the ``super-luminous spiral galaxies'' catalogued by \citet{Ogle+2016,Ogle+2019a}, except that, on average, they are slightly less massive. 
The reason for this is that interferometric \hi\ observations are easily obtainable only in the nearby Universe, with RFI severely undermining the quality of data at $z>0.05$.
Thus, the volume of space readily accessible in \hi\ is much smaller than that of the superspirals found in optical surveys up to $z\simeq0.3$. 
We expect that it will not be possible to enlarge our current sample significantly until the advent of the Square Kilometer Array (SKA).
Throughout this paper, we adopted homogenized distances and uncertainties from the Extragalactic Distance Database \citep[EDD,][]{Tully+2009} and the CosmicFlows project \citep[][]{Tully+2016}, when available.
Alternatively, we used median distances from the literature as reported in the NASA Extragalactic Database (NED).
\autoref{tab:sample} summarizes the main properties of the \galnum\ galaxies in our final sample. 
An atlas of galaxies, including optical images, \hi\ column-density maps and velocity fields, can be found in Appendix~\ref{app:kinmods}.

\subsection{Additional kinematical data}
\label{sec:ifudata}

Besides \hi\ data, for 6 galaxies, we also took advantage of publicly available kinematical data from Integral Field Unit (IFU) observations. 
Optical recombination lines from the warm ionized gas are observed at a much higher spatial resolution than the \hi\ line (although at lower spectral resolution) and are excellent to derive the inner rising part of the rotation curve of a galaxy and to constrain the contribution to the circular speed of the stellar component (see Section~\ref{sec:decomp}).
For this work, we found archival IFU data from the Mapping Nearby Galaxies at Apache Point Observatory survey  \citep[MaNGA,][]{Bundy+2015} for NGC2599, NGC2713 and UGC08179 and data from the Calar Alto Legacy Integral Field Area Survey \citep[CALIFA,][]{Sanchez+2012} for NGC1167, NGC1324 and NGC5635. 
In particular, we used datacubes from MaNGA DR7 and low spectral resolution datacubes (V500, channel width of 2\ang) from CALIFA DR3.

To isolate gas emission lines from IFU datacubes, we fitted and subtracted the stellar continuum. 
We first performed a Voronoi binning to achieve a $\mathrm{SNR} > 10$ per bin on the continuum \citep{Cappellari+2003}. The continuum was fitted in each bin with the Penalized PiXel-Fitting software \citep{Cappellari+2004} using MaStar Simple Stellar Population templates \citep[MASTARSSP,][]{Maraston+2020}. The best-fit stellar continuum was then subtracted from the spectrum in each spatial pixel. 
Finally, from the continuum-subtracted datacubes, we isolated the strongest emission lines for the kinematical modelling, i.e.\ the \ha\ line at $6564.63 \ang$ and of the \nii\ doublet at $6549.86-6585.27 \ang$.

\begin{table}
\centering
\caption{Properties of massive spiral galaxies analysed in this work. Columns: (1) Primary name; (2) Source of \hi\ data, either new data, WHISP, from \citet{Spekkens+2006} (SG06) or from \citet{Hunter+2013} (H13); 3) Assumed distance (EDD or NED); 4) Stellar mass from $W1$ WISE aperture photometry (typical error is 0.2 dex); 5) Mass of cold gas from \hi\ data, $M_\mathrm{gas}=1.36M_\hi$ (typical error is 0.1 dex).}
\label{tab:sample}
\begin{tabular}{lcccc}
\noalign{\vspace{1pt}}\hline\hline\noalign{\vspace{1pt}}
Name & \hi\ data & $D$ & $\log \frac{M_\star}{\rm M_\odot}$ & $\log \frac{M_\mathrm{gas}}{\rm M_\odot}$ \\
 &  & Mpc &  &  \\
(1)  & (2) & (3) & (4) & (5) \\

\noalign{\vspace{1pt}}\hline\noalign{\vspace{1pt}}
NGC0338 & WHISP & $78\pm11$ & 11.2 & 10.3 \\
NGC1167$^2$ & WHISP & $73\pm7$ & 11.3 & 10.3 \\
NGC1324$^2$ & SG06 & $72\pm10$ & 11.2 & 10.2\\
NGC2599$^1$ & WHISP & $73\pm8$ & 11.1 & 10.1\\
NGC2713$^1$ & new & $68\pm9$ & 11.4 & 10.2\\
NGC2862 & SG06 & $68\pm12$ & 11.1 & 10.2\\
NGC5440 & new & $57\pm4$ & 11.1 & 10.5 \\ 
NGC5533 & WHISP & $45\pm6$ & 11.0 & 10.4\\
NGC5635$^2$ & new & $72\pm13$ & 11.1 & 10.2\\
NGC5790 & new & $164\pm15$ & 11.3 & 10.3 \\
UGC02849 & SG06 & $138\pm25$ & 11.2 & 10.4\\
UGC02885 & H13 & $71\pm9$ & 11.2 & 10.6\\
UGC08179$^1$ & new & $232\pm18$ & 11.6 & 10.8\\
UGC12591 & new & $97\pm7$ & 11.7 & 10.0\\
UGC12811 & new & $174\pm13$ & 11.4 & 10.3\\
\noalign{\vspace{1pt}}\hline
\noalign{\vspace{2pt}}
\multicolumn{5}{l}{$^1$ These galaxies also have MaNGA data.}\\
\multicolumn{5}{l}{$^2$ These galaxies also have CALIFA data.}\\
\end{tabular}
\end{table}

\begin{figure*}
    \centering
    \includegraphics[width=\textwidth]{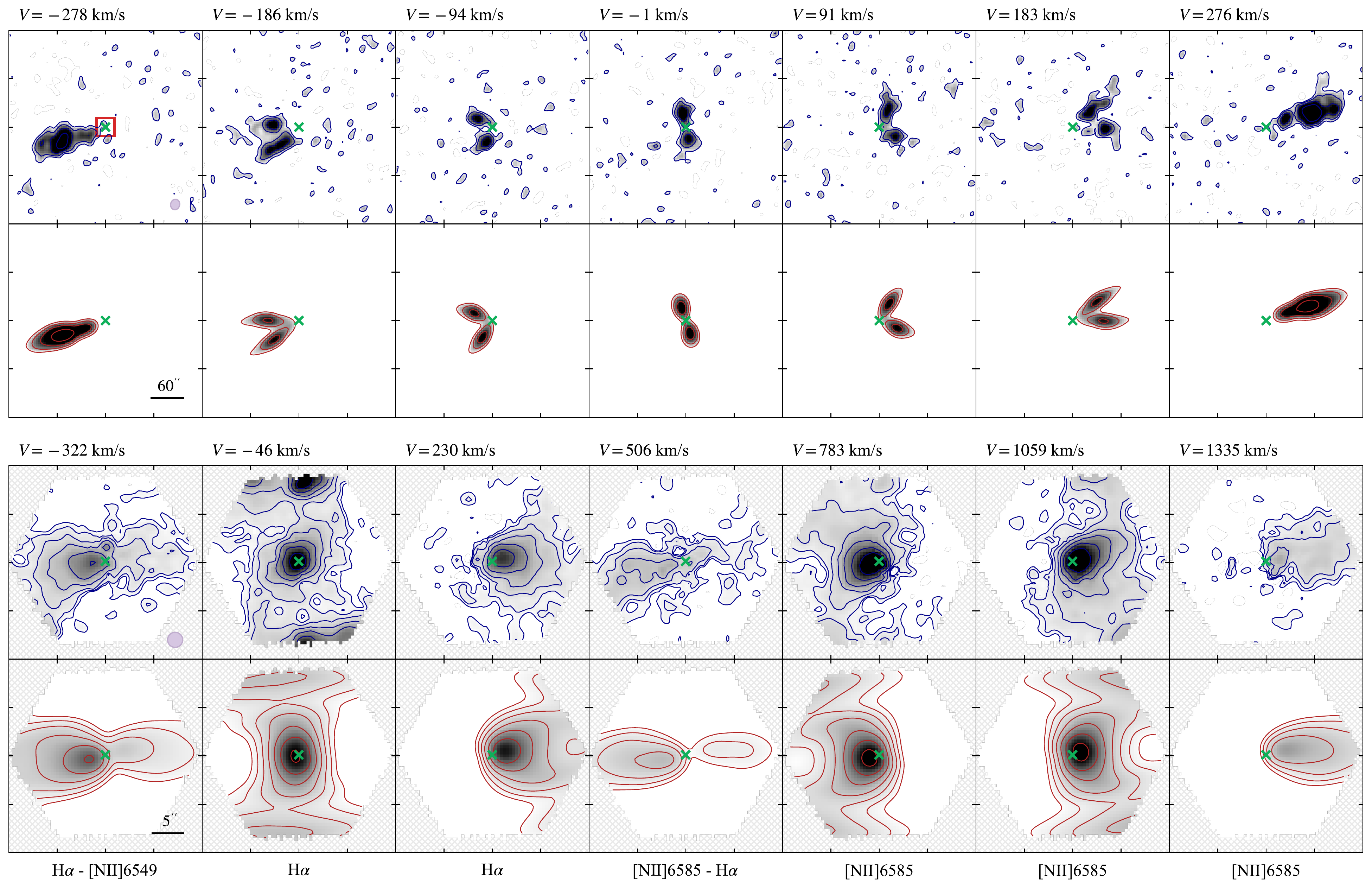}
    \caption{Comparison between observations (blue contours) and model (red contours) for the galaxy NGC2713. Top two rows show VLA \hi\ data, while bottom two rows are MaNGA \ha-\nii\ data of the inner regions of the galaxy. 
    On the first VLA channel map, a red rectangle denote the field of view of MaNGA data. 
    The green cross in each panel identifies the galaxy center. Contour levels are at $2.5\sigma_\mathrm{rms}\times2^n$, with $n = 0,1,...,6$. The beam is shown in the first channel map of both VLA and MaNGA data. For the MaNGA data, the first four channel maps include emission of the \ha\ line, while the last four channel maps include the emission of the \nii\ at 6585.27$\ang$. In the right part of the first channel, emission from the \nii\ at 6549.86$\ang$ is also visible. Velocity labels in the MaNGA data are centred on the systemic velocity of the \ha\ line.}
    \label{fig:modex}
\end{figure*}

\section{Methods}
\label{sec:methods}

\subsection{Photometry and baryonic masses}
\label{sec:phot}
Deep multi-band images from publicly available surveys were used to derive stellar surface-density profiles and to estimate stellar masses of galaxies in our sample. 
In particular, for 5 galaxies (NGC1167, NGC1324, NGC5533, UGC02849 and UGC02885), we used $z$-band images from the Panoramic Survey Telescope and Rapid Response System Survey DR2 \citep[Pan-STARRS1,][]{Chambers+2016}, while for the remaining 10 galaxies we used $z$-band images from the Dark Energy Spectroscopic Instrument (DESI) Legacy Surveys DR9 \citep{Dey+2019}. 
In addition, infrared $W1$-band images at 3.4 $\upmu$m from the Wide-field Infrared Survey Explorer \citep[WISE,][]{Wright+2010} were used to estimate the total stellar mass. 

As in the analysis performed in \citet{DiTeodoro+2021}, we derived stellar surface-brightness profiles from sky-subtracted $z$-band images through the \textsc{photoutils} package \citep{larry_bradley_2020_4044744} implemented within the \textsc{astropy} Python library \citep{astropy:2013,astropy:2018}. 
The light distribution of each galaxy was fitted with a set of elliptical annuli, each described by a centre $(x_0,y_0)$, a position angle $\phi$  and an ellipticity $\epsilon\equiv1-b/a$, where $b/a$ is the ratio of the ellipse's minor axis $b$ to the major axis $a$. 
The average surface-brightness $\Sigma_\star(R)$ was computed in each best-fit annulus.
Final errors for surface-brightness profiles were calculated by  adding in quadrature the rms variation along each annulus and the error on sky determination.
Derived profiles for all galaxies can be found in Appendix~\ref{app:kinmods}.

Stellar masses were estimated from background-subtracted 3.4$\upmu$m  images, where dust obscuration is expected to be negligible. Following \citet{Jarrett+2019}, we fitted a double Sérsic function to radial profiles and extrapolated it out to 3$R_\mathrm{d}$, where $R_\mathrm{d}$ is the exponential scale length of the disc. 
The total $W1$-band flux within 3$R_\mathrm{d}$ was then $K$-corrected 
and stellar mass was calculated from the total $W1$ luminosity assuming a constant mass-to-light ratio $\Upsilon_{W1}=0.6$ with a typical uncertainty of $\sim40\%$ \citep[e.g.,][]{Norris+2014,Rock+2015}.
The fourth column of \autoref{tab:sample} lists derived stellar masses $\mstar$ from $W1$ band, with typical errors of 0.2 dex, dominated by the uncertainty on the mass-to-light ratio. 

Finally, the atomic hydrogen mass for each galaxy was calculated directly from the total integrated \hi\ flux \citep{Roberts+1975}:

\begin{equation}
    M_\hi [\mo] = 2.36 \, D^2 \, \sum_{i=1}^N S_i \delta v  
\end{equation}

\noindent where $D$ is the distance in Mpc and $S_i \delta v $ is the total flux in the $i$-th velocity channel expressed in Jy $\kms$. 
This equation assumes that \hi\ gas is optically thin.
The total baryonic mass (stars + cold gas), ignoring contributions from ionized and molecular gas, is $M_\mathrm{bar} = \mstar + 1.36\, M_\hi$, where the factor 1.36 is to take into account the primordial abundance of Helium.

\subsection{3D tilted-ring kinematical modelling}
\label{sec:bbmod}
We modelled emission-line data with the kinematical fitting code \bba\ \citep{DiTeodoro&Fraternali15}\footnote{In particular, for this work we used non-stable version 1.6.3, which can be downloaded at \href{https://github.com/editeodoro/Bbarolo}{github.com/editeodoro/Bbarolo}.}. 
\bba\ builds 3D tilted-ring models of rotating disc galaxies and fits them directly to emission-line datacubes, taking into account well-known observational biases due to the finite spatial and spectral resolution of a telescope (e.g., beam smearing). 
This approach does not require the derivation of velocity maps and it proved to be very effective with a variety of emission-line data \citep[e.g.,][]{Loiacono+2019,BewketuBelete+2021,Ponomareva+2021,Dye+2022}. 
\bba\ describes a galaxy as a set of concentric, rotating rings. The main quantities that characterize each ring are: 1) the centre coordinates $(x_0,y_0)$; 2) the inclination angle $i$ with respect to the plane of the sky ($90\de$ for an edge-on discs); 3) the position angle $\phi$ of the projected major axis, defined anticlockwise from the North direction; 4) the systemic recession velocity \vsys; 5) the rotation velocity \vrot; 6) the intrinsic gas velocity dispersion \vdisp. All these parameters but the galaxy center and the systemic velocity were allowed to vary ring by ring during the kinematical modelling.

We used the same methodology in the kinematical analysis of radio and optical data. 
The modelling was performed on the \hi\ line and, when available, on the \ha-\nii\ lines simultaneously \citep[e.g.,][]{DiTeodoro+2018}. 
The kinematic centre was assumed to coincide with the photometric centre derived through the surface-density photometry (see Section~\ref{sec:phot}) and fixed to this value throughout the entire procedure.
Initial values for all other parameters were estimated automatically by \bba\ from the 0th and 1st moment maps.
A mask for the fit was created by using the source-finding masking algorithm with a primary threshold of $5\times\sigma_\mathrm{rms}$ and a secondary threshold of $3\times\sigma_\mathrm{rms}$, applied to datacubes smoothed by a factor of 1.5 (we refer to \bba's documentation and main paper for further details on these algorithms).
An azimuthal normalization scheme was adopted during the fit and, as a consequence, all models are fully symmetric in gas surface density and kinematics.

The fit was performed using a classical two-step approach. 
In the first step, all parameters were allowed to vary during the fit. 
Systemic velocity, inclination and position angles were then regularized and kept fixed during the second fitting step.
In particular, $\vsys$ was always fixed to a constant value, while $\phi$ and $i$ were regularized with either a constant, a polynomial function or a Bezier curve, depending on the results of the first fitting step. 
Finally, a second fit was performed with only the rotation velocity and velocity dispersion kept free. 
In both fitting steps, we used a $(\cos\theta)^2$ weight, where $\theta$ is the azimuthal angle ($\theta=0$ for major axis), to give more importance to pixels lying near the major axis of a galaxy.
Errors on all parameters were calculated through the default Monte Carlo method implemented in \bba, which basically samples the variation of the parameter space around the minimum of the residuals \citep[see][for details]{DiTeodoro&Fraternali15}.
Beside best-fit kinematical parameters, \bba\ also returns the \hi\ gas surface density $\Sigma_\mathrm{HI}$ averaged along the best-fit rings, which will be used for the mass modelling (Section~\ref{sec:decomp}) and calculation of the gas specific angular momentum (Section~\ref{sec:scaling}).

As an example, in \autoref{fig:modex} we show the best-fit 3D kinematical model for the galaxy NGC2713. For this particular galaxy, we had both new extended VLA \hi\ data (top rows) and IFU \ha-\nii\ data from MaNGA (bottom rows) for the very inner regions of the galaxy (red box in the uppermost left panel). 
A selection of channel maps of the data (blue contours) are compared to the corresponding channel maps of the best-fit model (red contours). 
For all velocity channels, the 3D model can accurately reproduce the line emission in both radio and optical data, signalling an optimal fit.
The resulting hybrid \hi-\ha\ rotation curve for NGC2713 can be found in the left panel of \autoref{fig:decomp}, where dots denote \hi\ and stars denote \ha-\nii.
In Appendix~\ref{app:kinmods}, we show a comparison between best-fit models and data for all galaxies through position-velocity slices taken along the apparent major and minor axes of the disc.
Most galaxies in our sample have very regular and symmetrical kinematics. 
We also checked for non-circular motions using the procedure of \citet[][]{DiTeodoro+2021b}, but we did not find any clear evidence of systematic streaming motions throughout these discs (with the exception of NGC5533, which has a long tidal tail in the external regions). 
This regularity allowed us to derive robust kinematical models and parameters for most galaxies.
 Notable exceptions are NGC5635 and UGC12591. 
The former has an asymmetric rotation curve and, on the approaching side, shows a significant amount of lagging extra-planar gas, which might indicate a recent accretion event \citep{Saglia+1988}, also supported by diffuse stellar tails in the optical images.
The latter is an extremely massive, lenticular galaxy with relatively little gas, and it is barely detected in our new VLA \hi\ data ($\mathrm{SNR}\lesssim3$ in most channels). 
Although we decided to model these two galaxies and we keep them in our sample, we note that their kinematics, and thus the following dynamical analysis, may be very uncertain, in particular for UGC12591.

\subsection{Rotation curve decomposition and mass models}
\label{sec:decomp}
Rotation curves derived in Section~\ref{sec:bbmod} were decomposed into the contribution of the different mass components. 
Given that our sample is made up by extremely massive galaxies, for which $\vrot\gg\vdisp$, we neglected any asymmetric drift correction \citep[e.g.][]{Bureau+2002} and we assumed $\vrot(R) \equiv \vcirc(R)$, where \vcirc\ is the circular velocity, i.e. the azimuthal component of the velocity induced in the equatorial plane by an axially symmetric gravitational potential $\phi(R,z)$.
The circular velocity profile can be written in terms of the matter components of a galaxy:

\begin{equation}
    \label{eq:vcirc}
    V^2_\mathrm{c}(R) = - R\,\frac{\partial \phi (R,z)}{\partial R} \bigg\rvert_{z=0} = 
    V^2_\mathrm{gas}(R) + V^{2}_\star(R) + V^2_\mathrm{DM}(R)
\end{equation}

\noindent where $V_\mathrm{gas}$ and $V_\star$ are the circular velocity induced by the gaseous and stellar discs, respectively, and $V_\mathrm{DM}$ is the circular velocity of the DM halo. 
The stellar circular velocity can be written as $V^2_\star = \Upsilon_\star V^2_{\star, 1}$, where $\Upsilon_\star$ is the mass-to-light ratio and $V_{\star, 1}$ is the circular velocity obtained under the assumption that $\Upsilon_\star=1$.
In other words, $\Upsilon_\star$ normalizes the contribution of the stellar component and is needed because we measure the distribution of light rather than mass.
Usually, the gas component $V_\mathrm{gas}$ and the shape of the stellar component $V_{\star,1}$ can be determined from the data, while the DM component $V_\mathrm{DM}$ and the mass-to-light ratio $\Upsilon_\star$ need to be fitted to the observed rotation curve of a galaxy.

The gravitational potential, thus the circular speed through Eq.~\ref{eq:vcirc}, of the stellar and gaseous components can be directly calculated from the observed mass distributions via numerical integration \citep[see e.g.,][]{Casertano+1983,Cuddeford+1993}.
We assumed that both stars and gas lie in discs with a classical vertical profile for an iso-thermal, self-gravitating layer, given by a squared hyperbolic secant:

\begin{equation}
    \rho_\mathrm{d}(R,z) = \frac{\Sigma_\mathrm{d}(R)}{2z_\mathrm{0}}  \mathrm{sech}^2 \left( \frac{z}{z_\mathrm{0}} \right)
\end{equation}

\noindent where $\Sigma_\mathrm{d}(R)$ is the deprojected (i.e.\ face-on) radial profile for a given component, i.e.\ $\Sigma_\star$ (Section~\ref{sec:phot}) for stars and $\Sigma_\mathrm{gas} = 1.36\, \Sigma_\hi$ (Section~\ref{sec:bbmod}) for gas.
Contributions from molecular and ionized gas are expected to be negligible, especially in the outer regions of a galaxy, and are usually not computed in rotation curve decompositions.
For both the stellar and the gaseous disc, we assume a constant scale height of $z_0=300$ pc.
We note that, for massive galaxies like those in our sample, the choice of a different vertical profile and/or scale height would have no significant effect on the rotation curve decomposition \citep{ManceraPina+2022}.
Finally, for five galaxies with a clear bulge component in the rotation curve and optical images (NGC1167, NGC1324, NGC2713, NGC5635 and UGC08179), we also used a spheroidal stellar component in addition to the disc component, derived from a bulge-disc decomposition of the surface-brightness profiles. 

\begin{figure*}
    \centering
    \includegraphics[width=\textwidth]{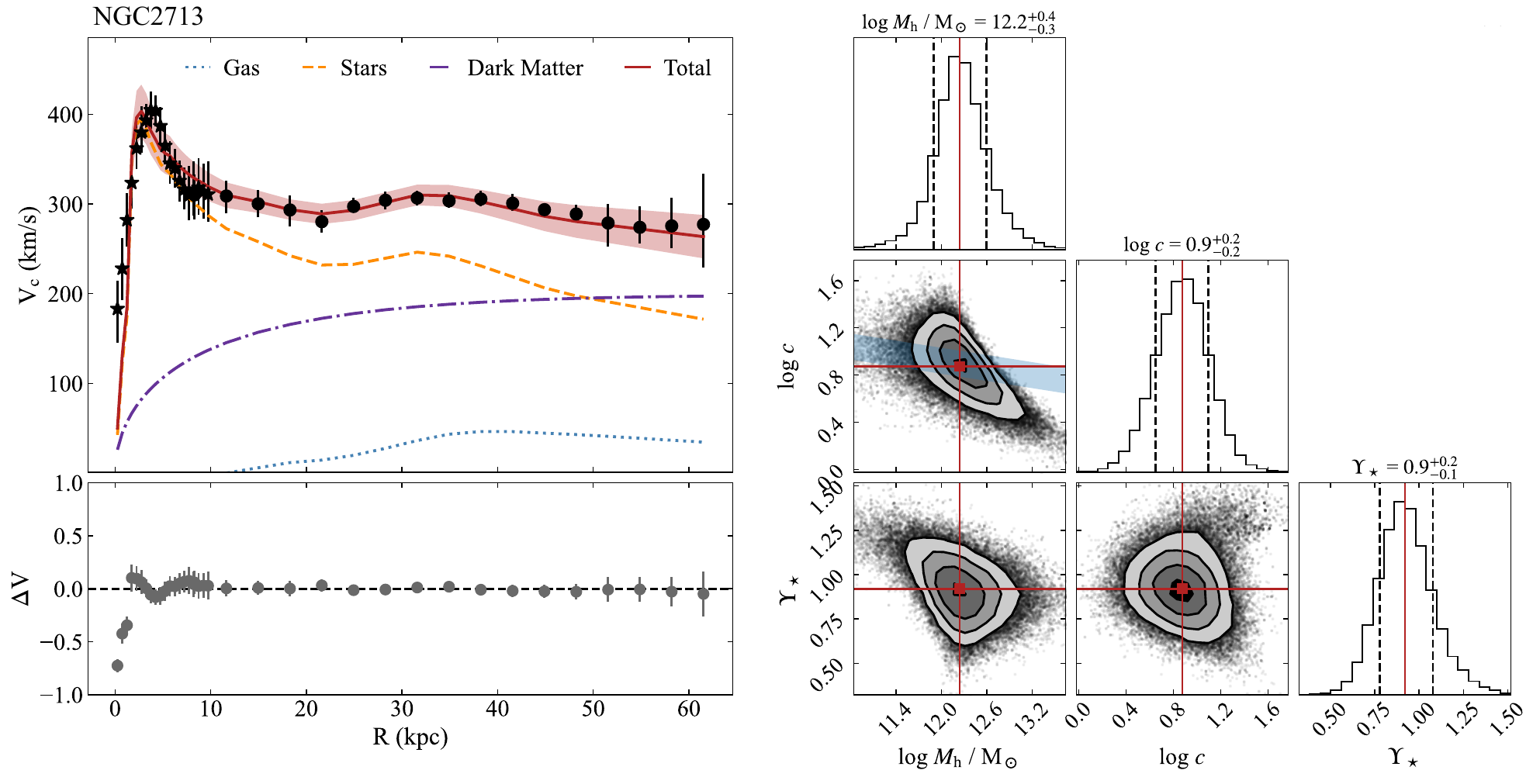}
    \caption{Rotation curve decomposition for the galaxy NGC2713. 
    \emph{Left:} Mass modelling of the hybrid \hi-\ha\ rotation curve (\hi=dots, \ha=stars). 
    Mass components are gas (blue-dotted line), stars (bulge and disc, dashed orange line) and dark matter (dashed-dotted purple line). 
    The red-solid line represents the total circular velocity of the model. 
    Bottom panel shows the fractional difference between the modelled and the measured circular velocity curve, $\Delta V = (V_\mathrm{tot}-V_\mathrm{obs})/V_\mathrm{obs}$. 
    \emph{Right:} Triangle plot of the MCMC sampling. 2D posterior distributions are shown as contour plots, with contour drawn at 1$\sigma$, 2$\sigma$ and 3$\sigma$ confidence levels. Histograms are the 1D posterior distributions. Full red lines denote the 50th percentile values, dashed black lines are the 15.87th and 84.13th percentiles. The blue band represents the $c_{200}-M_{200}$ relation by \citet{Dutton+2014} used as a prior. }
    \label{fig:decomp}
\end{figure*}

The dark matter distribution is modelled as a \citet*{Navarro+1996} (NFW) spherical halo profile:

\begin{equation}
    \label{eq:NFW}
    \rho_\mathrm{DM}(r) = \frac{4\rho_\mathrm{s}}{(r/r_\mathrm{s})(1+r/r_\mathrm{s})^2}
\end{equation}

\noindent where $r=\sqrt{R^2+z^2}$ is the spherical radius, $r_\mathrm{s}$ is a scale radius and $\rho_\mathrm{s}$ is the density at $r_\mathrm{s}$.
The characteristic density $\rho_\mathrm{s}$ is given by:

\begin{equation}
    \label{eq:rhos}
    \rho_\mathrm{s} = \frac{M_{200}}{16\pi r^3_\mathrm{s} [\ln(1+c_{200})-c_{200}/(1+c_{200})]}
\end{equation}

\noindent where $M_{200}$ is the DM mass within a radius $R_{200}$, where the average density is 200 times the critical density of the Universe, and $c_{200} \equiv R_{200}/r_\mathrm{s}$ is the dimensionless concentration parameter. 
Cosmological DM simulations indicate that $M_{200}$ anti-correlates with $c_{200}$ \citep{Dutton+2014,Ludlow+2014}. 
For brevity, in the rest of the paper, we will use the notation \mhalo\ and $c$ to indicate $M_{200}$ and $c_{200}$, respectively.

Given the above parametrization, our rotation curve decomposition procedure has only three free parameters: $\mathbf{x}=(\Upsilon_\star, \mhalo, c)$.
For each galaxy in our sample, we found the best-fit $\Upsilon_\star$, \mhalo\ and $c$ via a Monte-Carlo Markov-Chain (MCMC) sampling, using the Python implementation \texttt{emcee} \citep{Foreman-Mackey+2013}.
We used a standard $\chi^2$ likelihood function $\mathcal{P}$, defined as:

\begin{equation}
    \chi^2 = -\ln \mathcal{P} (V_\mathrm{rot}  \, | \, V_\mathrm{mod}(\mathbf{x})) = \sum_{i=1}^N \left[ \frac{\vrot(R_i)-V_\mathrm{mod}(R_i)}{\Delta_i}  \right]^2 
\end{equation}

\noindent where $\vrot(R_i)$ and $V_\mathrm{mod}(R_i)$ are the observed and model circular velocity at the $i$-th radius $R_i$, respectively, and $\Delta_i$ are the errors on the rotation velocity.

Appropriate priors on all free parameters were set during the MCMC sampling.
With regard to $\Upsilon_\star$, 
spiral galaxies are known to have a range of mass-to-light ratios that can depend on galaxy morphological type \citep{Bell+2001,Portinari+2004,Fall+2013}, due to their different star formation histories and stellar populations.
Testing different Gaussian priors for $\Upsilon_\star$, in a range of values predicted by stellar population models, we found that this has little effect on the resulting mass modeling of different galaxies. 
For simplicity, for all galaxies we therefore used a Gaussian prior with mean 0.8 and standard deviation 0.3, which broadly embraces values expected in the $z$-band \citep[e.g.,][]{Meidt+2014,McGaugh+2014}.
The prior on $\mhalo$ is, instead, flat over a wide range $6 < \log \mhalo/\mo \leq 15$.
Finally, for the $c$ parameter, we used a lognormal prior with mean given by the $c_{200}-M_{200}$ relation and standard deviation of 0.11 dex around the relation derived from $\Lambda$CDM simulations \cite[equation 4 of][]{Dutton+2014}\footnote{For simplicity and consistency with previous work in this field, we adopt the $c_{200}-M_{200}$ relation in $\Lambda$CDM simulations without baryons. 
Inflows of baryons within dark halos generally increase their concentration, while outflows decrease it.  
Thus, the net effect of these flows on the $c_{200}-M_{200}$ relation likely depends on the cooling rates in the protogalactic gas as well as the prescriptions for star formation, black hole growth, and stellar and AGN feedback -- a situation that is difficult to quantify reliably. 
In some recent hydrodynamical simulations (\textsc{Eagle}) the alteration of the $c_{200}-M_{200}$ relation by baryons is negligible \citep{Schaller+2015}, while in others (\textsc{Illustris}) it is significant \citep{Chua+2017}. 
}.
We stress that a strong non-uniform prior on the concentration parameter is necessary to infer reasonable constraints on the halo parameters \citep[see e.g.,][]{Katz+2017,Posti+2019}.
We sampled posterior probability distributions with 100 walkers running 10000 steps, including a warm-up phase of 1000 steps. 
For each free parameter, we assumed the 50th percentile value of the 1D posterior distribution as central value, while lower and upper errors were taken at the 15.87th and 84.13th percentiles, corresponding to the standard deviation for a Gaussian distribution.

\autoref{fig:decomp} illustrates the rotation curve decomposition for our example galaxy NGC2713.
The right panel shows a triangle plot with 1D joint posterior distributions and 2D marginalized posterior distributions.
Central best-fit values for $\mhalo$, $c$ and $\Upsilon_\star$ are shown as red-solid lines, 1$\sigma$ quantiles as black dashed lines on the 1D histograms.
The posterior distributions indicate a well-attained sampling and converged chains.
The resulting mass model and rotation curve decomposition is illustrated in the left panel. 
The calculated gas and stellar velocity curves are shown as dotted-blue and dashed-orange lines, respectively, while the fitted DM velocity curve is the dashed-dotted purple line. 
The total circular velocity curve of the model is the solid-red line, compared to the observed rotation curve (black dots and stars) derived as described in Section~\ref{sec:bbmod}. 
Note the bump in the stellar circular velocity around 30 kpc in radius, caused by a similar feature in the stellar light profile, which corresponds to a bump in the rotation curve \citep[ ``Renzo's rule",][]{Sancisi+2004}.
This rotation curve decomposition returned a DM halo mass for NGC2713 of $\log (\mhalo/\mo) \simeq 12.2\pm0.3$. 

We obtained similarly good mass models for most galaxies in our sample. Rotation curve decompositions and MCMC corner plots for all galaxies are shown again in the atlas of Appendix~\ref{app:kinmods}. 
Thanks to the extended \hi\ rotation curves, we were able to put robust constraints on the dark matter halo masses of 13 out of \galnum\ galaxies. 
Again, the most problematic galaxies are NGC5635 and UGC12591, for which the kinematic and mass modelling did not allow us to unequivocally decompose the contribution of the visible and dark matter. 

The galaxies in our sample all lack publicily-available CO observations, preventing us from including molecular gas in our calculations of baryonic quantities. 
However, we expect the contribution of molecules to be relatively minor, based on observations of other massive spirals, which typically have $M_\mathrm{mol} \sim \ M_\hi \sim 0.1\mstar$, mostly concentrated toward the galactic centres \citep{Saintonge+2022}. 
This will have little effect on the outer rotation curves, and thus on the derived mass of the halo and angular momentum of the disc. 
Finally, because the stellar component should always dominate the circular velocity in the inner regions, we do not expect any significant effect on the estimated total stellar mass either.

\begin{figure*}
    \centering
    \includegraphics[width=\textwidth]{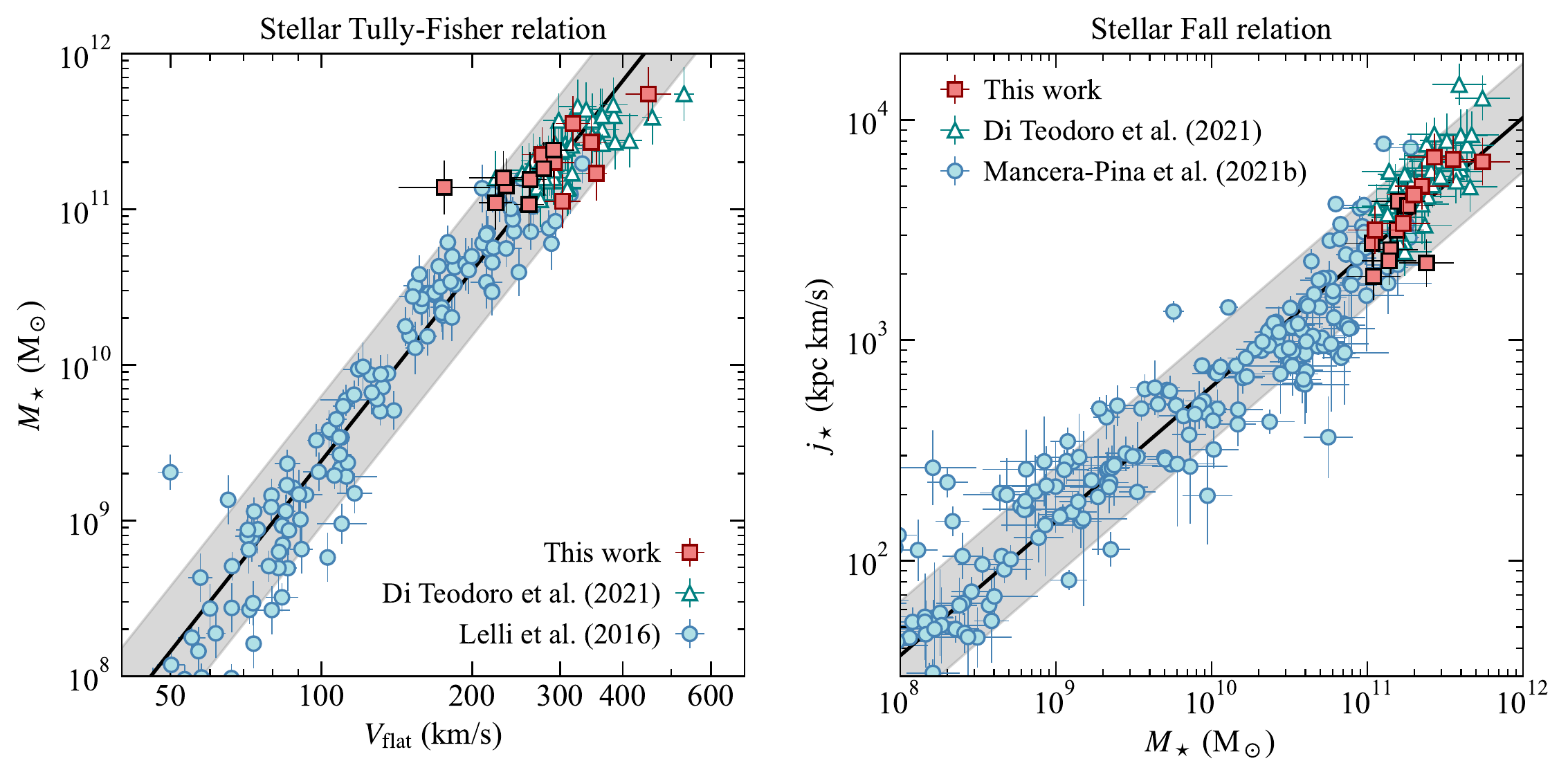}
    \vspace{2pt}
    \includegraphics[width=\textwidth]{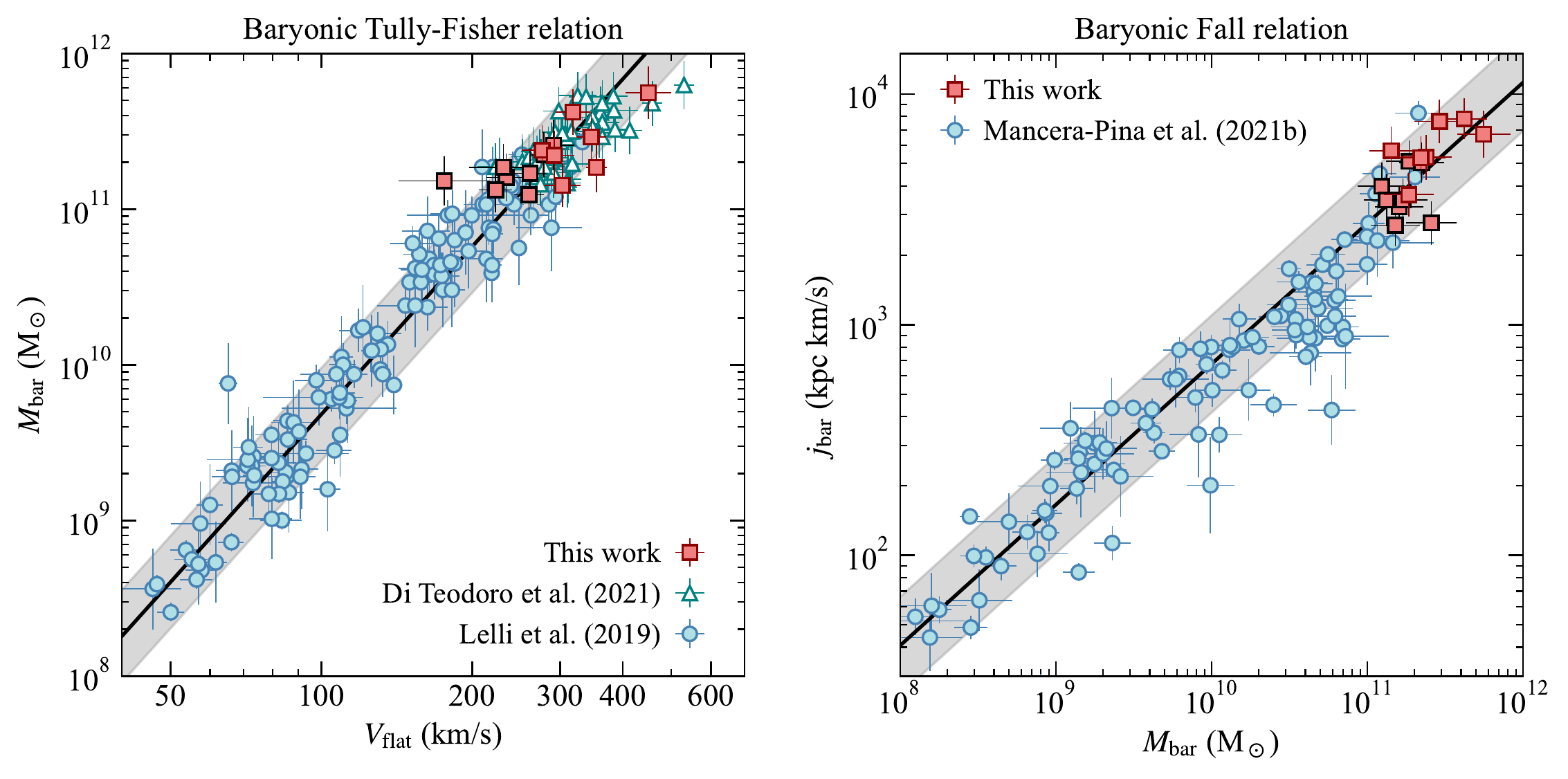}
    \caption{Tully-Fisher relations (left panels) and Fall relations (right panels) for our sample of extremely massive spiral galaxies with extended \hi\ rotation curves (red squares). 
    Markers of galaxies with new VLA data are shown with red edges, while archival galaxies with black edges.  
    We also plot galaxies with measurements from \hi\ rotation curves from the literature \citep[blue circles,][]{Lelli+2016,Lelli+2019,ManceraPina+2021b}. 
    Open green triangles denote the high-mass sample of \citet{DiTeodoro+2021}, with measurements coming from long-slit \ha\ rotation curves only. 
    Black lines and grey shaded regions indicate best-fit relations and orthogonal scatter around them, respectively. 
    The best-fit parameters and $1\sigma$ errors of these relations are listed in \autoref{tab:bestfit}.
    }
    \label{fig:scaling}
\end{figure*}

\section{Results}
\label{sec:results}

\subsection{Tully-Fisher and Fall scaling relations}
\label{sec:scaling}
We placed our massive spiral galaxies on the two most important scaling relations for discs: the \citet{Tully+1977} relation, between mass and rotation velocity, and the \citet{Fall+1983} relation, between mass and specific angular momentum $j=J/M$. 
We explored both the stellar and baryonic (stars + atomic gas) versions of these relations.

For the Tully-Fisher relations, we used the average velocity along the flat part of the rotation curve $\vflat$, which has been shown to minimize the scatter of the relation \citep{Verheijen+2001,Lelli+2019}.
This characteristic speed was computed following the iterative algorithm of \citet{Lelli+2016b} (see their equations 1 and 2), i.e. by requiring that the rotation velocities are flat within $5\%$ over at least the last three measured points. 
We note that all our galaxies have very flat rotation curves in the outer regions and many more points are usually included in the computation of \vflat. 
For consistency with the literature, we used stellar masses derived from the $W1$-band WISE photometry with a fixed $\Upsilon_{W1}=0.6$ (Section~\ref{sec:phot}) rather than those derived from the mass modelling. 
The stellar masses obtained by these two different methods never differ more than the typical uncertainty of 0.1-0.2 dex (see Appendix~\ref{app:mstar}). 
For the baryonic mass, we simply assume $M_\mathrm{bar}=\mstar+1.36M_\hi$.

The left panels of \autoref{fig:scaling} show the stellar (top) and baryonic (bottom) Tully-Fisher relations.
Together with our sample (red squares), we plot less massive galaxies from the SPARC sample \citep{Lelli+2016,Lelli+2019} (blue circles) and the higher-$z$ sample ($0.04\lesssim z\lesssim 0.28$) of superspiral galaxies by \citet{DiTeodoro+2021} (open triangles).
The latter only have \ha\ long-slit rotation curves, which typically do not extend further out than the optical disc, while \hi\ rotation curves for most galaxies in this work reach radii $1.5-2$ times larger than the optical disc.
Throughout this paper, all datapoints derived from extended \hi\ rotation curves (red and blue symbols), in all the relations presented, were determined by similar analysis techniques and with similar assumptions.
In both the stellar and baryonic Tully-Fisher relations, our galaxies lie at the high-mass end of a single, unbroken power-law relation.
We fit linear functions in the form $\log M = \beta \log \vflat + \log B$ to the Tully-Fisher relations in \autoref{fig:scaling}, using an orthogonal distance regression method to take into account errors on both axes.
The best-fit parameters are $(\beta,\log B) = (4.06\pm0.14,1.26\pm0.32)$ for the stellar relation and $(\beta,\log B) = (3.58\pm0.10,2.52\pm0.22)$ for the baryonic relation (black lines).
These values are in good agreement with previous determinations of the relations at lower masses in the local Universe \citep[e.g.][]{McGaugh12,Zaritsky+2014,Lelli+2016b,Lelli+2019,Papastergis+2016,Ponomareva+2017,Ponomareva+2021}.
The best-fit values and 1$\sigma$ errors for all relations investigated in this paper are summarized in \autoref{tab:bestfit}.
Our new analysis therefore confirms the recent finding by \citet{DiTeodoro+2021} that the Tully-Fisher relations extend at least up to $\log M/\mo \simeq 11.7$ as unbroken power laws
.

Our sample includes the S0/Sa galaxy UGC12591, which has been known for decades as the fastest rotating disc in the local Universe \citep{Giovanelli+1986}.
Contrary to recent results by \citet{Ray+2022}, we find that this galaxy is consistent with the Tully-Fisher relation within the uncertainties. 
This discrepancy is ascribable to the fact that \citeauthor{Ray+2022} estimated a significantly lower stellar mass ($\log \mstar / \mo \simeq$ 11.2 vs our 11.7) and they used the \ha\ maximum rotation velocity from \citet{Giovanelli+1986}, which exceeds 500 $\kms$ in the inner regions, while we use the flat part of the extended \hi\ rotation curve ($\vflat\simeq450 \, \kms$). 
We note that our stellar mass estimate is in agreement with other measurements in the literature \citep[e.g.,][]{Dai+2012,Li+2017} and that our \hi\ data, although noisy, seem to disfavour a flat velocity substantially larger than 450 $\kms$.

\begin{figure*}
    \centering
    \includegraphics[width=\textwidth]{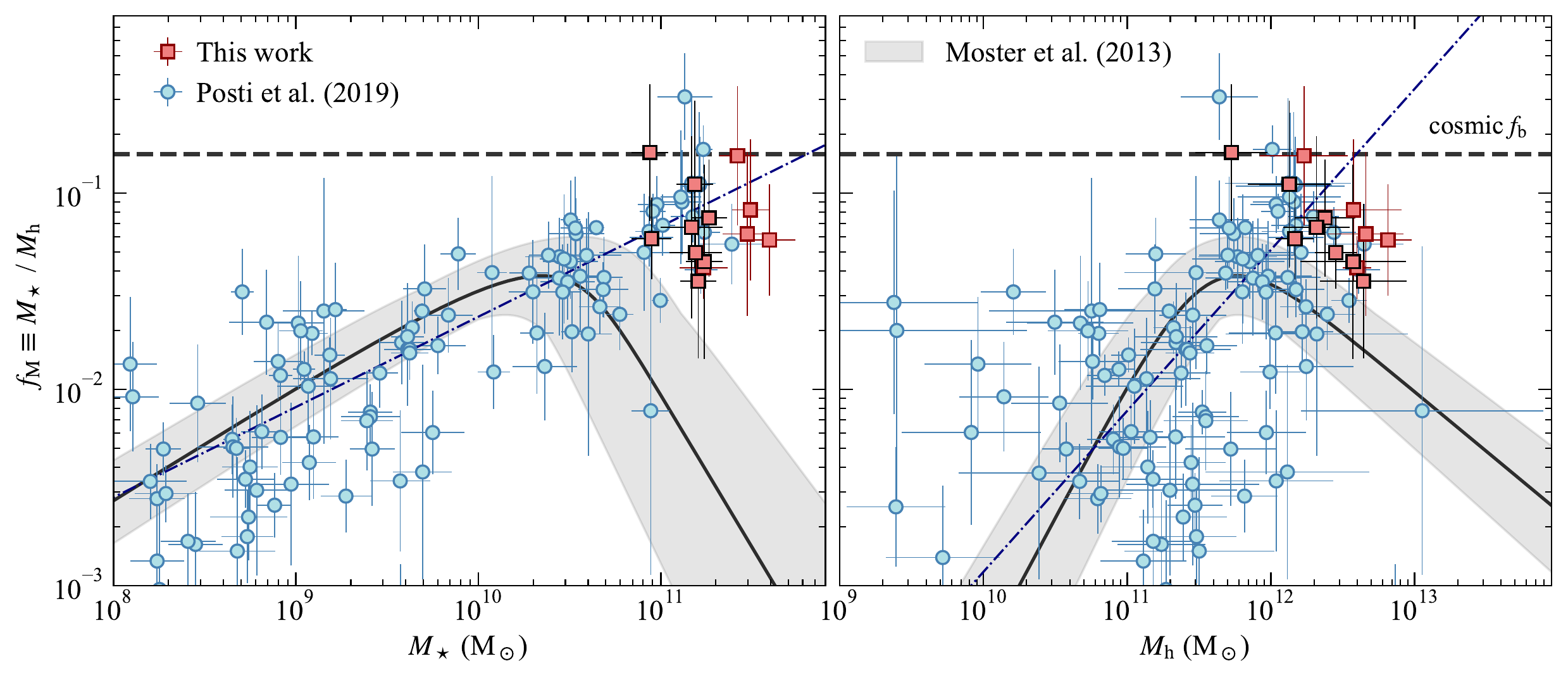}
    \caption{Stellar-to-halo mass relations $f_\mathrm{M}-\mstar$ (left) and $f_\mathrm{M}-\mhalo$ (right). In both panels, we plot our galaxy sample (red squares) and the spiral galaxy sample by \citet{Posti+2019b}, obtained with an identical technique and the same priors. As in \autoref{fig:scaling}, galaxies with archival data have black edges. 
    The horizontal black-dashed lines denote the cosmic baryon fraction, i.e.\ $f_\mathrm{M} = f_\mathrm{b} = 0.156$. 
    The black full curves are the mean SHMR from abundance matching by \citet{Moster+2013}, with grey bands representing 1$\sigma$ deviations. 
    The diagonal blue dashed-dotted lines are best-fit linear relations to the entire galaxy sample.
    The best-fit parameters and $1\sigma$ errors of these relations are listed in \autoref{tab:bestfit}.
    }
    \label{fig:shmr}
\end{figure*}

\begin{figure}
    \centering
    \includegraphics[width=0.49\textwidth]{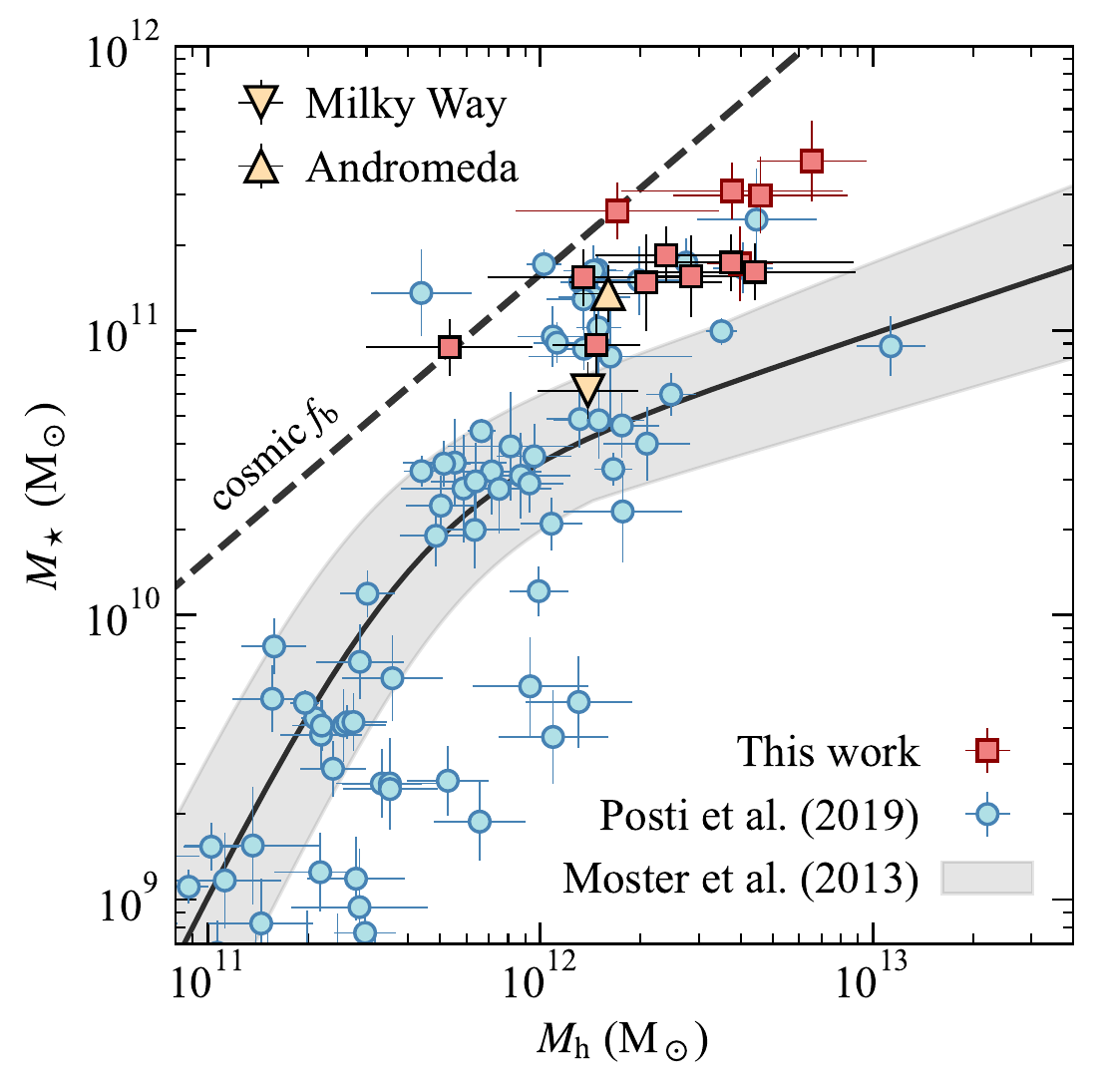}
    \caption{Stellar-to-halo mass relation $\mstar - \mhalo$. Symbols as in \autoref{fig:shmr}. In this plot we also show the Milky Way \citep[downfacing triangle,][]{Posti+2019c} and the Andromeda galaxy \citep[upfacing triangle,][]{Corbelli+2010}.}
    \label{fig:shmr2}
\end{figure}

For the Fall relations, the specific angular momentum $j_k$ for a given component $k$ (stars, $\jstar$, or gas, $j_\mathrm{gas}$) inside a radius $R$ can be calculated as:

\begin{equation}
\label{eq:jstar}
    j_k(<R) = \frac{J_k(<R)}{M_k(<R)} = \frac{\int_0^R \, \Sigma_k(R') \, {R'}^2 \, V_{k,\mathrm{rot}}(R') \,  dR' }{\int_0^R \, \Sigma_k(R') \, R' \, dR' }
\end{equation}

\noindent 

\noindent where $\Sigma_k$ and $V_{k,\mathrm{rot}}$ are the mass surface density and the azimuthal velocity of either the gas ($\Sigma_\mathrm{gas}$, $V_\mathrm{gas,rot}$) or the stars ($\sigmastar$, $V_\mathrm{\star,rot}$). 
For gas, $V_\mathrm{gas,rot}$ is simply the \hi\ rotation curve $\vrot$, while for stars we also assume co-rotation with the gas, i.e.\ $V_\mathrm{\star,rot}=\vrot$.
Although stars usually have larger velocity dispersions and thus rotate slower than cold gas, this velocity lag is expected to be negligible in high-mass disc galaxies \citep{Obreschkow+2014,Posti+2018b}.
For the total baryons, $j_\mathrm{bar}$ was calculated by using $\Sigma_\mathrm{bar} = \Sigma_\star + 1.36\Sigma_\mathrm{\hi}$ and $V_\mathrm{bar} = \vrot$.
Equation~\ref{eq:jstar} was used to compute the specific angular momentum of a given matter component as a function of radius.
The total values of $\jstar$ and $j_\mathrm{bar}$ for each galaxy were then taken from the converged cumulative $j$ profiles \citep[for details, see][]{Posti+2018b,DiTeodoro+2021}.

The right panels of \autoref{fig:scaling} show the stellar (top) and baryonic (bottom) Fall relations.
In addition to our sample (red squares), we plot galaxies from \citet{ManceraPina+2021} (blue circles) and the higher-$z$ \ha\ sample from \citet{DiTeodoro+2021} (green triangles, only in the stellar relation). 
The black lines are linear fits in the form $\log j = \alpha \log M + \log A$, with best-fit values $(\alpha,\log A) = (0.61\pm0.08,-3.31\pm0.26)$ for the stellar relation and $(\alpha,\log A) = (0.61\pm0.07, -3.27\pm0.24)$ for the baryonic relation, consistent with previous studies \citep[e.g.][]{Fall+1983,Fall+2013,Fall+2018,Posti+2018b,Murugeshan+19,ManceraPina+2021,ManceraPina+2021b,DiTeodoro+2021,Hardwick+2022}.
As with the Tully-Fisher relation, the baryonic form of the Fall relation is tighter than the stellar form, with an orthogonal scatter $\sigma_\perp$ of the datapoints around the best-fit power law decreasing from 0.21 dex to 0.18 dex.
At the high-mass end, galaxies tend to lie slightly above these best-fit relations. 
This might be a sign that the Fall relation has a weak bend at high masses, or it may be a bias due to our preference for selecting fast-rotating galaxies (see Section~\ref{sec:HIdata}).
However, both the Bayesian Information Criterion (BIC) and the Akaike Information Criterion (AIC) disfavour a double power-law model over a single power-law model.
We note finally that our slope $\alpha\simeq0.61$ for discs is very close to the value of 2/3 for DM halos in a $\Lambda$CDM cosmology, thus already indicating that the specific angular momentum ratio $f_\mathrm{j}$ is nearly independent of mass.

\begin{table}
\centering
\caption{Best-fit parameters $\gamma$ and $\log C$ for relations in the form $\log (y) = \gamma\log (x) + \log C$. For the Tully-Fisher and Fall relations, we also list the orthogonal perpendicular scatter $\sigma_\perp$ around the best-fit relation.}
\label{tab:bestfit}
\begin{tabular}{lrrc}
\noalign{\vspace{1pt}}\hline\hline\noalign{\vspace{1pt}}
Relation ($y-x$) & $\gamma$\hspace{17pt} & $\log C$\hspace{11pt} & $\sigma_\perp$\\

\noalign{\vspace{1pt}}\hline\noalign{\vspace{1pt}}
\textit{Tully-Fisher relations} \\
$\mstar-\vflat$ & $4.06\pm0.14$ & $1.26\pm0.32$ & 0.10\\
$M_\mathrm{bar}-\vflat$ & $3.58\pm0.10$ & $2.52\pm0.22$ & 0.07 \\
\noalign{\vspace{1pt}}\hline
\textit{Fall relations} \\
$\jstar-\mstar$ & $0.61\pm0.08$ & $-3.31\pm0.26$ & 0.21 \\
$j_\mathrm{bar}-M_\mathrm{bar}$ & $0.61\pm0.07$ & $-3.27\pm0.24$ & 0.18 \\
\noalign{\vspace{1pt}}\hline
\multicolumn{2}{l}{\textit{Stellar-to-halo mass relations}}\\
$f_\mathrm{M}-\mstar$ & $0.46\pm0.08$ & $-6.23\pm0.38$ & - \\
$f_\mathrm{M}-M_\mathrm{h}$ & $0.82\pm0.16$ & $-11.13\pm1.72$ & - \\
$\mstar-M_\mathrm{h}$ & $0.59\pm0.06$ & $5.74\pm0.32$ & - \\
\noalign{\vspace{1pt}}\hline
\multicolumn{2}{l}{\textit{Retained specific angular momentum relations}}\\
$\fj-\mstar$ & $0.16\pm0.07$ & $-1.70\pm0.39$ & - \\
$\fj-M_\mathrm{h}$ & $0.13\pm0.09$ & $-1.62\pm0.46$& -  \\
$f_\mathrm{j,bar}-M_\mathrm{bar}$ & $0.04\pm0.03$ & $-0.14\pm0.39$ & - \\
$f_\mathrm{j,bar}-M_\mathrm{h}$ & $-0.02\pm0.04$ & $-1.62\pm0.48$ & - \\
\noalign{\vspace{1pt}}\hline
\end{tabular}
\end{table}

\subsection{Stellar-to-halo mass relations}
\label{sec:dm}

In this Section, we present the SHMRs for our sample of massive spiral galaxies. 
\autoref{fig:shmr} shows the SHMR in the form $f_\mathrm{M} \equiv \mstar/\mhalo$ as a function of $\mstar$ (left) and as a function of $\mhalo$ (right).
The galaxies analysed in this work are represented by red squares.
The two galaxies for which we could not obtain a good DM halo constraint (NGC5635 and UGC12591) are not shown in this and the following figures. 
We also plot the lower-mass disc galaxy sample of \citet{Posti+2019b}, derived with an identical mass decomposition procedure, as blue circles.
The horizontal thick dashed line highlights the cosmic baryon fraction, i.e.\ $f_\mathrm{M} = f_\mathrm{b}$ = 0.156.
The grey band shows the SHMR estimated by \citet{Moster+2013} using abundance matching, which represents the standard SHMR in the field \citep{Wechsler+2018}. 
We note that other SHMRs in the literature \citep[e.g.,][]{Rodriguez-Puebla+2015,Mandelbaum+2016,Behroozi+2019} have qualitatively similar shapes to that of \citet{Moster+2013}.
In both panels, the diagonal blue dotted-dashed line denotes a simple linear fit to the datapoints. 
\autoref{fig:shmr2} shows instead the high-mass end of the $\mstar - \mhalo$ relation, using the same symbols as \autoref{fig:shmr}. 
This represents a good alternative visualization of the SHMR because errors on the plotted quantities are not correlated by construction.
In the $\mstar - \mhalo$ diagram, as a reference, we also plot the positions of the Milky Way from \citet{Posti+2019c} and M31 from \citet{Corbelli+2010}.

The difference between abundance matching and our dynamical fits in the high-mass regime of the SHMR is evident, particularly from the $f_\mathrm{M}-\mstar$ and the $\mhalo - \mstar$ relations.
Our galaxy sample confirms a picture where $f_\mathrm{M}$ increases monotonically with $\mstar$, with no indication of a peak in the range $10.5\leq \log \mstar / \mo \leq 11 $, where the SHMR derived from abundance matching turns over.
For example, for a galaxy with $\mstar = 2\times10^{11} \, \mo$, the \citet{Moster+2013} SHMR implies $f_\mathrm{M}\simeq 0.04$ and $\mhalo \simeq 6\times10^{13} \, \mo $, while with our linear relation we find $f_\mathrm{M}\simeq 0.09$ and $\mhalo \simeq 2\times10^{12} \, \mo$.
In other words, massive spirals inhabit less massive halos than galaxies in general, which tend to be lenticulars and ellipticals at high mass.
These findings agree with and improve upon other recent studies of dark matter halos in late-type galaxies at slightly lower masses \citep[$\mstar$ up to $1-2 \times 10^{11}\,\mo$,][]{Posti+2019b,Li+20}.

Finally, we note that this difference holds regardless of the assumptions made during the mass modeling procedure. 
In particular, we tested the effect of using different DM halo profiles beside the NFW profile of Equation~\ref{eq:NFW}, including pseudo-isothermal, cored \citet{Burkert+1995} and \citet{Einasto+1965} profiles. 
We also varied the centre of the prior on the mass-to-light ratio in the $z$-band over the range $0.3\lesssim\Upsilon_\star\lesssim1.3$ indicated by stellar population synthesis models \citep[e.g.,][]{McGaugh+2014,Schombert+2019}. 
We found that, although changing the above assumptions may marginally modify our DM halo mass estimates, the overall picture of the SHMR shown in \autoref{fig:shmr} does not change significantly \citep[see also Appendix A2 in][for similar tests on the SHMR]{Posti+2019b}.

\begin{figure*}
    \centering
    \includegraphics[width=\textwidth]{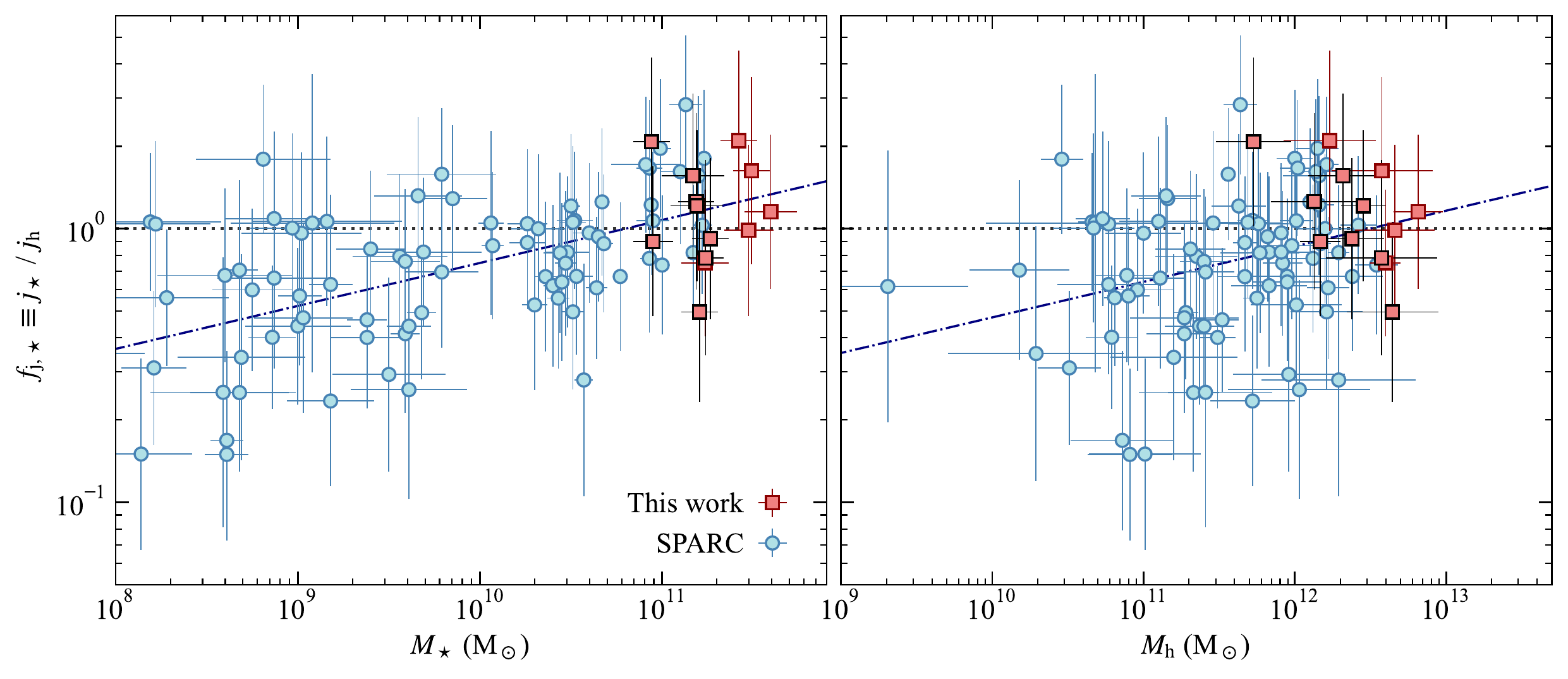}
    \vspace{2pt}
    \includegraphics[width=\textwidth]{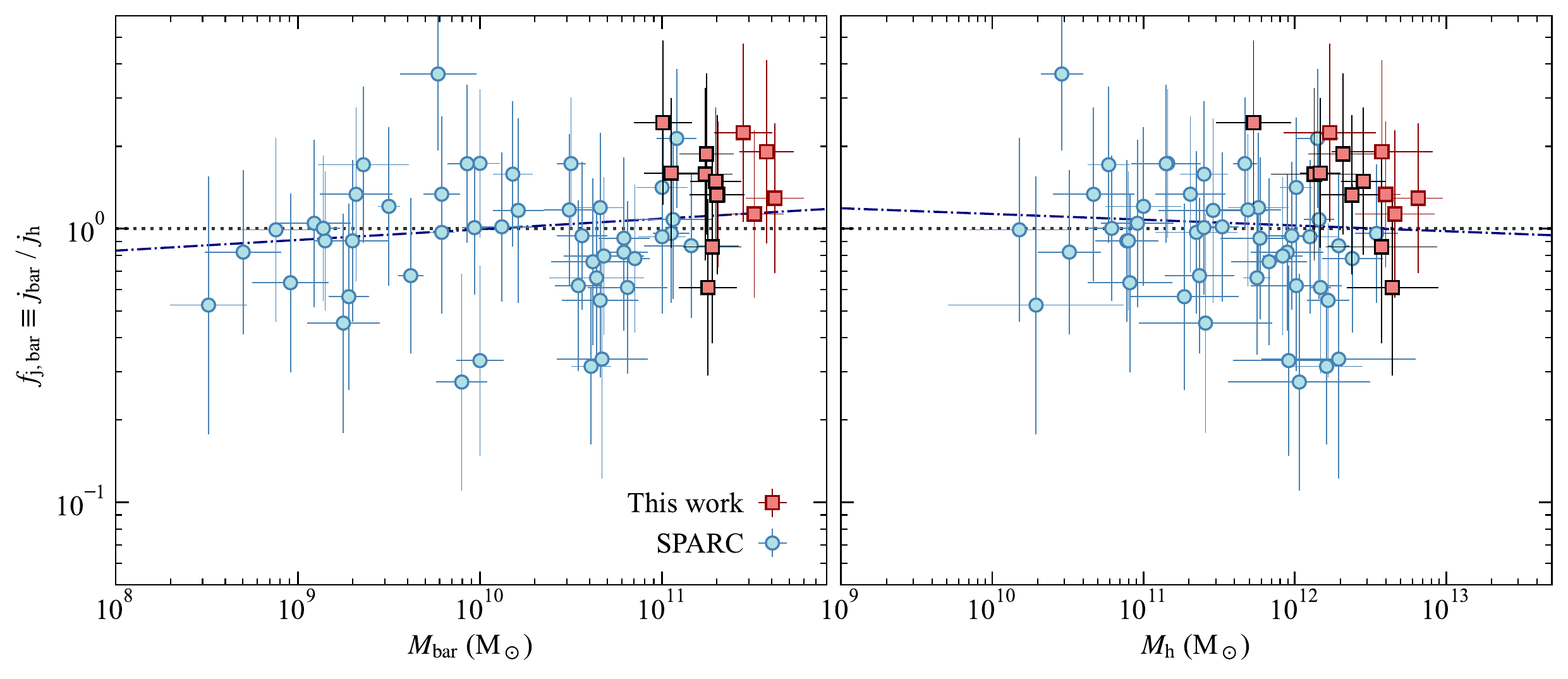}
    \caption{Retained fraction of angular momentum of stars ($\fj\equiv\jstar/j_\mathrm{h}$, upper panels) and baryons ($f_\mathrm{j,bar}\equiv j_\mathrm{bar}/j_\mathrm{h}$, lower panels) as a function of stellar/baryonic mass (left) and DM halo mass (right).
    We plot our galaxy sample (red squares, archival data with black edges) and the SPARC sample with measurements from \citet{Posti+2018b}, \citet{Posti+2019b} and \citet{ManceraPina+2021b}. 
    The horizontal black dotted lines indicate $f_\mathrm{j}=1.$
    The tilted blue dashed-dotted lines are best-fit linear relations to the entire galaxy sample. 
    The best-fit parameters and $1\sigma$ errors of these relations are given in \autoref{tab:bestfit}.
    }
    \label{fig:fj}
\end{figure*}

\subsection{Angular momentum retention}
\label{sec:dmj}
As a final step, we studied how the specific angular momentum of stars $\jstar$ and baryons $j_\mathrm{bar}$ relates to the specific angular momentum of DM halos $j_\mathrm{h}$ through the ratios $\fj \equiv \jstar / j_\mathrm{h}$ and $f_\mathrm{j,bar} \equiv j_\mathrm{bar} / j_\mathrm{h}$, also known as ``retained fractions of angular momentum''.
In particular, for each galaxy, we estimated the average value of $f_\mathrm{j,k}$ for a given component $k$ (stars or baryons) as

\begin{equation}
\label{eq:fj}
\begin{aligned}
        \left< f_\mathrm{j,k} \right> = j_\mathrm{k} \left< \frac{1}{j_\mathrm{h}} \right> &= j_\mathrm{k} \, \frac{\int_{0}^{+\infty} \, [j_\mathrm{h}(\lambda)]^{-1} \, p(\lambda) \, d\lambda}
    {\int_{0}^{+\infty} \, p(\lambda) \, d\lambda } \approx \\
    &\approx \frac{j_\mathrm{k}}{\sqrt{2}R_\mathrm{h}V_\mathrm{h}} \left< \frac{1}{\lambda} \right> \approx 32.5 \, \frac{j_\mathrm{k}}{\sqrt{2}R_\mathrm{h}V_\mathrm{h}} \hspace{10pt} .
\end{aligned}
\end{equation}

\noindent Here, $R_\mathrm{h}=R_{200}$ and $V_\mathrm{h}=V_{200}$ are the virial radius and circular velocity of the DM halo, where its mean enclosed density is 200 times the cosmic critical density, and we have used the \citet[][]{Bullock+2001} approximate relation $j_\mathrm{h}(\lambda) \approx \sqrt 2 \lambda R_\mathrm{h}V_\mathrm{h}$ between the spin parameter $\lambda$ and specific angular momentum $j_\mathrm{h}$ of the halo.
The distribution of spin parameters $p(\lambda)$ for halos in $\Lambda$CDM is known to be very close to a log-normal distribution with mean $\log \lambda = -1.456$ and $\sigma_{\log \lambda} = 0.22$ dex, independent of halo mass \citep{Bullock+2001,Bett+2007}. 
This implies $\left< 1/ \lambda \right> \simeq 32.5$.
Errors on $f_\mathrm{j,k}$ were calculated by bootstrapping, i.e.\ taking the standard deviation over 10k realizations of $f_\mathrm{j,k}$ obtained by randomly sampling $j_\mathrm{k}$, $R_\mathrm{h}$, $V_\mathrm{h}$ and $\lambda$ according to their distributions.
We remark that the calculation of $f_\mathrm{j}$ is analogous to that of $f_\mathrm{M}$.
In both ratios, the numerator (\mstar\ or $j_\mathrm{k}$) is determined from observations, while the denominator ($\mhalo$ or $j_\mathrm{h}$) is inferred by imposing physically-motived priors, i.e.\ the lognormal priors on the concentration $c$ for $\mhalo$ (Section~\ref{sec:decomp}) and on the spin parameter $\lambda$ for $j_\mathrm{h}$.

The upper panels of \autoref{fig:fj} show how $\fj$ varies as a function of $\mstar$ (left) and $\mhalo$ (right), while the lower panels display $f_\mathrm{j,bar}$ as a function of $M_\mathrm{bar}$ (left) and $\mhalo$ (right).
As in previous plots, the red squares represent the 13 massive galaxies analysed in this work for which we could obtain robust estimates of DM halo masses. 
The blue circles represent SPARC galaxies, for which we cross-matched $\jstar$ and/or $j_\mathrm{bar}$ from \citet{Posti+2018b} and \citet{ManceraPina+2021b} to $j_\mathrm{h}$ calculated from rotation curve decomposition by \citet{Posti+2019b}. 
The tilted blue dotted-dashed lines represent linear fits to the relations.
Although the datapoints have significant errors and scatter, $\fj$ increases only marginally with stellar and halo mass.
This dependence is very weak, with best-fit slopes of 0.16 (vs $\mstar$) and 0.13 (vs $\mhalo$).
The baryonic ratio $f_\mathrm{j,bar}$ is even flatter, basically constant with mass.
The mean and median of $\fj$ across the entire spiral sample are 0.70 and 0.75, while those of $f_\mathrm{j,bar}$ are 0.93 and 0.96, respectively.
This $\sim$20\% offset between the specific angular momentum of stars and baryons can be understood in terms of gas-to-stellar ratios of specific angular momentum and mass. 
Because, on average, $j_\mathrm{gas}\simeq2\jstar$ \citep{ManceraPina+2021} and $M_\mathrm{gas}\simeq0.1\mstar$, the gas adds approximately 20\% to the total specific angular momentum, accounting for the difference between $\fj$ and $f_\mathrm{j,bar}$. 
Moreover, because low mass galaxies have in general larger gas fractions than more massive galaxies, the relative contribution of $j_\mathrm{gas}$ decreases with mass, causing the flattening of the baryonic relation with respect to the stellar relation.
Our galaxies sit at the high-mass end of the relations and show $f_\mathrm{j,bar}$ (and $\fj$) values consistent with unity, i.e.\ $\jstar \simeq j_\mathrm{bar} \simeq j_\mathrm{h}$.
In other words, galactic discs have retained essentially all the initial specific angular momentum they acquired from tidal torques, with little or no dependence on stellar and halo mass.
Our findings agree with other recent estimates of $\fj$ in spiral galaxies \citep{Fall+2013,Huang+2017,Posti+2019,DiTeodoro+2021}.


The most important quantities estimated in this work are summarized in \autoref{tab:dynquant}.

\subsection{Constraining the galaxy--halo connection with the observed scaling relations}

In the $\Lambda$CDM cosmology, the formation of DM halos is relatively simple and well understood, leading to the scaling relations $\mhalo \propto V_\mathrm{h}^3$ and $j_\mathrm{h}  \propto \mhalo^{2/3}$ \citep*[e.g.,][]{MovdBW10}. In contrast, the formation of the baryonic components of galaxies by dissipative collapse, star formation, feedback, and other processes within the dark halos is much more complex and still under active investigation.  
The net effects of these formation processes are conveniently summarized by the stellar fractions $f_\mathrm{M} \equiv \mstar / \mhalo$, $f_\mathrm{j,\star} \equiv  \jstar / j_\mathrm{h}$, and $f_\mathrm{V} \equiv \vflat / V_\mathrm{h}$ and their baryonic counterparts. 
The stellar Tully-Fisher and Fall relations can then be rewritten in this notation as $\mstar \propto f_\mathrm{M} ( \vflat / f_\mathrm{V})^3$ and $j_\star \propto f_\mathrm{j,\star} ( \mstar / f_\mathrm{M,\star})^{2/3}$.

In the previous Section, we derived the (logarithmic) slopes of these linear scaling relations and the corresponding mass dependencies of the fractions $f_\mathrm{M}$ and $f_\mathrm{j,\star}$ assuming that the measured quantities $\mstar$, $\jstar$, and $\vflat$ are all independent of each other (with results listed in \autoref{tab:bestfit}).  
In the $\Lambda$CDM cosmology, however, $\mstar$, $\jstar$, and $\vflat$ are linked by the fractions $f_\mathrm{M}$, $f_\mathrm{j,\star}$, and $f_\mathrm{V}$ to the halo spin parameter $\lambda$ induced by tidal torques, which, as mentioned in Section~\ref{sec:dmj}, has a well-known, nearly log-normal probability distribution $p(\lambda)$, independent of halo mass $\mhalo$ and overdensity $\Delta$ \citep{Bullock+2001}. 
In a recent work, \citet{Posti+2019} have designed a novel statistical procedure that simultaneously fits for the dependencies of $f_\mathrm{M}$, $f_\mathrm{j,\star}$, and $f_\mathrm{V}$ on $\mstar$ or $\vflat$ while also enforcing the required $p(\lambda)$ as a prior. 
For a complete description of the method and results, we refer interested readers to the \citet{Posti+2019} paper.

\citet{Posti+2019} applied this procedure to a sample of 138 nearby galaxies with near-infrared images and \hi\ rotation curves \citep[mostly from][]{Lelli+2016b}, covering a wide range of stellar masses.  
We have repeated this analysis with the addition of the 15 high-mass spirals with new \hi\ rotation curves presented in this paper\footnote{We have adopted this approach in order to test whether the extension of the sample to higher masses alters the results significantly. We have also checked that the results are not sensitive to the exact definition of the sample at lower masses.}.
The results are $f_\mathrm{M} \propto \mstar^{0.31 \pm 0.08}$, $f_\mathrm{j,\star} \propto \mstar^{0.02 \pm 0.06}$, and $f_\mathrm{V} \propto \mstar^{-0.01 \pm 0.04}$ (for the linear model), i.e., a significant increase of $f_\mathrm{M}$ with $\mstar$ but no dependence of $f_\mathrm{j,\star}$ or $f_\mathrm{V}$ on $\mstar$, in excellent agreement with the findings of \citet[][]{Posti+2019} (see their Table B.1). 
These fits also agree reasonably well with those listed here in \autoref{tab:bestfit} given the different assumptions involved, i.e.\ one with independent $\mstar$, $\jstar$, and $\vflat$ and the other with these quantities linked by $p(\lambda)$. 
Thus, this formalism, which returns a nearly constant $\fj\sim0.8-1$ and $f_\mathrm{V}\sim1$, and a linearly increasing $f_\mathrm{M}$, reinforces our conclusions that the power-law forms of the Tully-Fisher and Fall relations descend naturally from the scaling relations of their parent DM halos.

\begin{table*}
\centering
\caption{Physical quantities derived through kinematical and dynamical modelling. Columns: (1) Primary name; (2) Velocity of the flat part of the rotation curve $\vflat$; (3) Specific angular momentum of stara $\jstar$; (4) Specific angular momentum of baryons $j_\mathrm{bar}$; (5) Stellar mass \mstar\ from rotation curve decomposition; (6) Dark-matter halo mass \mhalo; (7) Concentration parameter $c$; (7) Stellar-to-halo mass ratio $f_\mathrm{M}\equiv\mstar/\mhalo$; (8) Retained fraction of specific angular momentum of stars $\fj=\jstar/j_\mathrm{h}$; (9) Retained fraction of specific angular momentum of baryons $f_\mathrm{j,bar}=j_\mathrm{bar}/j_\mathrm{h}$.
Kinematical and/or mass decompositions for NGC5635 and UGC12591 are extremely uncertain. We do not use these two galaxies in the SHMRs and in the $f_\mathrm{j}-M$ relations.}
\label{tab:dynquant}
\begin{tabular}{lccccccccc}
\noalign{\vspace{1pt}}\hline\hline\noalign{\vspace{1pt}}
Name & $\vflat/\kms$ & $\log\frac{\jstar}{\mathrm{kpc}\, \kms}$ & $\log\frac{j_\mathrm{bar}}{\mathrm{kpc}\, \kms}$ & $\log \frac{\mstar}{\mo}$ & $\log \frac{\mhalo}{\mo}$ & $\log c$ & $f_\mathrm{M}$ & $\fj$ & $f_\mathrm{j,bar}$\\
(1)  & (2) & (3) & (4) & (5) & (6) & (7) & (8) & (9) & (10)\\

\noalign{\vspace{1pt}}\hline\noalign{\vspace{1pt}}
NGC0338 & $234\pm21$ & $3.41\pm0.11$ & $3.51\pm0.10$ & $11.2\pm0.1$ & $12.1\pm0.3$ & $0.9\pm0.1$ & $0.11\pm0.09$ & $1.26\pm0.94$ & $1.58\pm1.16$\\
NGC1167 & $291\pm28$ & $3.35\pm0.10$ & $3.44\pm0.10$ & $11.2\pm0.1$ & $12.7\pm0.3$ & $0.9\pm0.1$ & $0.04\pm0.02$ & $0.49\pm0.37$ & $0.61\pm0.45$\\
NGC1324 & $261\pm20$ & $3.50\pm0.09$ & $3.54\pm0.10$ & $11.2\pm0.1$ & $12.6\pm0.4$ & $0.9\pm0.1$ & $0.04\pm0.03$ & $0.78\pm0.64$ & $0.86\pm0.69$\\
NGC2599 & $176\pm36$ & $3.36\pm0.11$ & $3.43\pm0.11$ & $10.9\pm0.1$ & $11.7\pm0.2$ & $1.0\pm0.1$ & $0.16\pm0.12$ & $2.07\pm1.46$ & $2.44\pm1.70$\\
NGC2713 & $276\pm26$ & $3.70\pm0.11$ & $3.73\pm0.11$ & $11.5\pm0.1$ & $12.2\pm0.3$ & $0.9\pm0.2$ & $0.15\pm0.12$ & $2.10\pm1.59$ & $2.25\pm1.69$\\
NGC2862 & $260\pm18$ & $3.44\pm0.09$ & $3.60\pm0.10$ & $11.3\pm0.1$ & $12.4\pm0.2$ & $0.9\pm0.1$ & $0.07\pm0.05$ & $0.92\pm0.63$ & $1.33\pm0.89$\\
NGC5440 & $303\pm25$ & $3.50\pm0.08$ & $3.75\pm0.09$ & $11.2\pm0.1$ & $12.6\pm0.1$ & $1.0\pm0.1$ & $0.04\pm0.02$ & $0.75\pm0.46$ & $1.34\pm0.82$\\
NGC5533 & $223\pm17$ & $3.29\pm0.10$ & $3.54\pm0.10$ & $10.9\pm0.1$ & $12.2\pm0.1$ & $1.0\pm0.1$ & $0.06\pm0.03$ & $0.89\pm0.56$ & $1.59\pm1.01$\\
NGC5635 & $354\pm18$ & $3.53\pm0.11$ & $3.57\pm0.11$ & $11.2\pm0.1$ & $13.4\pm0.5$ & $0.8\pm0.1$ & $0.01\pm0.01$ & $0.25\pm0.22$ & $0.28\pm0.24$\\
NGC5790 & $292\pm27$ & $3.66\pm0.11$ & $3.72\pm0.10$ & $11.5\pm0.1$ & $12.7\pm0.3$ & $0.9\pm0.1$ & $0.06\pm0.05$ & $0.99\pm0.71$ & $1.13\pm0.80$\\
UGC02849 & $231\pm36$ & $3.63\pm0.10$ & $3.71\pm0.10$ & $11.2\pm0.2$ & $12.3\pm0.3$ & $1.0\pm0.1$ & $0.07\pm0.06$ & $1.56\pm1.08$ & $1.88\pm1.27$\\
UGC02885 & $278\pm16$ & $3.61\pm0.10$ & $3.70\pm0.10$ & $11.2\pm0.2$ & $12.5\pm0.2$ & $0.9\pm0.1$ & $0.05\pm0.02$ & $1.21\pm0.77$ & $1.49\pm0.93$\\
UGC08179 & $318\pm25$ & $3.82\pm0.11$ & $3.89\pm0.11$ & $11.4\pm0.1$ & $12.6\pm0.4$ & $0.9\pm0.1$ & $0.08\pm0.06$ & $1.63\pm1.27$ & $1.91\pm1.48$\\
UGC12591 & $450\pm47$ & $3.81\pm0.11$ & $3.82\pm0.11$ & $11.6\pm0.1$ & $14.0\pm0.8$ & $0.9\pm0.1$ & $0.01\pm0.01$ & $0.18\pm0.23$ & $0.19\pm0.24$\\
UGC12811 & $346\pm24$ & $3.83\pm0.11$ & $3.88\pm0.11$ & $11.6\pm0.2$ & $12.8\pm0.2$ & $0.9\pm0.1$ & $0.06\pm0.03$ & $1.15\pm0.74$ & $1.29\pm0.81$\\
\noalign{\vspace{1pt}}\hline
\noalign{\vspace{2pt}}
\end{tabular}
\end{table*}


\section{Discussion}
\label{sec:disc}
Our analysis clearly shows that massive discs sit on the continuation of the most important scaling relations for late-type galaxies (Tully-Fisher and Fall relations) and that they live in dark matter halos that scale almost linearly with stellar mass. 
This implies that disc galaxies are a self-similar population of objects closely related to their dark matter halos over a wide range of mass, i.e. high-mass discs are simply scaled up versions of low-mass discs.
Moreover, we find that galactic discs, on average, have almost as much specific angular momentum as their halos, i.e.\ they retain nearly all the initial angular momentum imparted by tidal torques.

We observe that massive discs have systematically lower halo masses, corresponding to higher stellar-to-halo mass ratios $f_\mathrm{M}$, than those expected from conventional abundance matching prescriptions \citep{Wechsler+2018}.
For example, our spiral galaxies typically have DM masses a factor $3-8$ smaller than those implied by the \citet{Moster+2013} relation.
A likely explanation for this discrepancy was recently put forward by \citet{PostiFall21}, who argued that the SHMR has a secondary correlation with galaxy type, i.e.\ the relation splits in two different branches at high masses, a rising one for late-type galaxies and a falling one for early-type galaxies. 
Abundance matching prescriptions are mostly based on the general galaxy population; however, the ratio of early-type to late-type galaxies varies significantly with stellar and halo mass. 
In particular, the galaxy population is heavily dominated by massive early-type spheroids for $M_\star\gtrsim5\times10^{10} \, \mo $ \citep{Kelvin+2014}. 
In this context, it is not surprising that massive discs, which are rare at high masses and not well represented by abundance matching prescriptions, are discrepant with respect to those predictions. 
Recent cosmological simulations of galaxy formation also find morphology-dependent SHMRs \citep{Correa+2020,Marasco+2020,Rodriguez-Gomez+2022}, although these differences are somewhat weaker than the empirically-based SHMRs presented here. 

Our findings further support this idea of a splitting at high-masses of the SHMR, which implies the existence of two distinct evolutionary pathways for building up massive discs and massive spheroids.
Massive spiral galaxies like the ones in our sample must have been 
evolving in near isolation, continuing to grow in mass and gradually transforming gas into stars, with their star formation mainly sustained through a relatively smooth accretion of gas from the surroundings and/or minor mergers, and with a regulation mechanism dictated by feedback from young stars.
Such gradual evolution would move galaxies up to the high-mass ends of the scaling laws and up to the rising branch of the SHMR, without introducing a prominent feature like a break or a bend. 
These galaxies somehow avoided disruptive events \citep[see also][]{Saburova+2018,Saburova+2021}, like major mergers and strong AGN feedback in the form of outflows and radiation, capable of drastically impeding gas inflows and star formation, hence reducing $f_\mathrm{M}$, and at the same time causing the dynamical heating of the stellar body and the growth of the spheroidal component \citep[e.g.,][]{Hopkins+2006,Hopkins+2010,Martin+2018}.
These events would have moved massive discs off the scaling relations and off the rising branch of the SHMR.
In simplified terms, discs mostly know about hierarchical assembly, which gives them their self-similarity, and are mainly regulated by gas accretion and star-formation feedback. 
They have relatively small spheroids, thus relatively small black holes \citep[e.g.,][]{Kormendy+1995}, and weak AGN feedback. 
Conversely, massive elliptical galaxies mostly know about major mergers and strong AGN feedback, which can drastically inhibit stellar growth and move these galaxies toward the falling branch of the SHMR \citep[see][for further discussion of this picture]{PostiFall21}.

Another interesting aspect of this evolutionary picture for massive spirals is their high overall efficiency in converting available baryons into stars. 
Galaxies in our sample typically have very high stellar fractions $\fstar = f_\mathrm{M}/f_\mathrm{b}\simeq 0.3-0.9$, often referred to as ``global star formation efficiencies".
In contrast, $L_\star$ galaxies ($\mhalo\simeq10^{12}\,\mo)$, which are usually considered to be at the maximum star formation efficiency \citep{Wechsler+2018}, typically only have $\fstar\simeq 0.2-0.3$. 
More massive elliptical and lenticular galaxies have even lower baryon  conversion efficiencies. 
Furthermore, the fraction of baryons in massive discs is also extremely high, compared to lower mass discs and to spheroids. 
If, in addition to the stellar fraction $\fstar$, we consider also the fraction of atomic gas $f_\mathrm{at}=1.36f_\mathrm{\hi}$, of molecular gas $f_\mathrm{mol}$ \citep[for example through the $M_\hi - M_\mathrm{H_2}$ relation,][]{Catinella+2018}, and of ionized gas in the hot CGM $f_\mathrm{ion}\simeq 0.1-0.3f_\mathrm{b}\mhalo$ \citep[e.g.,][]{Bregman+2018}, we obtain a total baryonic fraction $f_\mathrm{bar} = \fstar + f_\mathrm{at} + f_\mathrm{mol} + f_\mathrm{ion} \sim 1$ for most galaxies in our sample.
In other words, consistent with the findings by \citet{Posti+2019b}, we confirm that the most massive disc galaxies have remarkably little, if any, missing baryons.

Finally, our work indicates that baryons in discs retain a large fraction of the halo specific angular momentum, i.e.\ $f_\mathrm{j,bar}\simeq1$, independent of mass.
Such a behavior was already predicted by early models of disc formation on the basis of simple conservation arguments, starting with \citet{Fall+1980} and then adopted in essentially all analytical and semi-analytical models of galaxy formation \citep[see also][and many subsequent papers]{Dalcanton+1997,Mo+98}.
We do not yet have a full understanding of this noteworthy result ($f_\mathrm{j}\simeq1$) because there are several physical processes that can change the specific angular momentum of baryons over galactic lifetimes \citep[e.g.,][]{Romanowsky+2012,DeFelippis+2017}. 
We remark, however, that smooth inflow of gas from a gradually settling hot halo does promote the conservation of specific angular momentum \citep{Hafen+2022}. 
It is also worth noting that $\fstar\simeq 1$ for massive spiral galaxies makes the apparent conservation of $j_\mathrm{bar}$ less puzzling: if most baryons are in the disc, there is then little circumgalactic medium  in which to hide additional mass and angular momentum.

\section{Summary and conclusions}
\label{sec:conclusions}
In this paper, we studied the dark matter halos and the main scaling relations of some of the most massive spiral galaxies in the local Universe. 
These discs have stellar masses $\mstar>10^{11} \, \mo$, are relatively gas-rich and have typical outer rotation velocities that approach or exceed 300 $\kms$. 
This is probably the most complete sample of very massive spiral galaxies with \hi\ interferometric data that will be available for the next decade.
Taking advantage of our new VLA \hi\ data, together with other archival \hi\ data, we derived extended rotation curves through 3D kinematical modelling for a sample of \galnum\ massive galaxies. 
We then obtained constraints on their dark matter halos by using a MCMC approach to decompose their rotation curves into the contribution of stars, cold gas and dark matter halo.
We used the results from our dynamical modelling to build the two most important scaling relations for discs, namely the Tully-Fisher relation ($M-V$) and the Fall relation ($j-M$), in both their stellar and baryonic versions. 
Finally, we derived the stellar-to-halo mass relation (SHMR, e.g. $\mstar$ vs $\mhalo$) and we investigated how the stellar and baryonic-to-halo specific angular momentum ratios vary as functions of stellar and halo masses.

Our investigation reinforces recent findings that massive discs lie on the high-mass ends of both the Tully-Fisher and the Fall relations. 
These relations appear to be unbroken power laws spanning roughly 5 orders of magnitude in mass, from dwarf galaxies ($\mstar\sim10^7 \, \mo$) to the most massive spiral galaxies ($\mstar\sim10^{12} \, \mo$).
We also show that the ratio $f_\mathrm{M}\equiv\mstar/\mhalo$ increases monotonically with mass and does not turn over above $\mstar\simeq5\times10^{10}$, in sharp contrast with widely-used SHMRs from abundance matching.
Finally, we showed that the ratio between specific angular momentum of baryons and halos ($f_\mathrm{j, bar} \equiv j_\mathrm{bar}/j_\mathrm{h}$) is close to unity not only for massive spiral galaxies, but for discs in general, with little or no dependence on galaxy mass.
This is a remarkable vindication of the basic premise, adopted in early models of galaxy formation, that galactic discs have approximately the same specific angular momentum as their surrounding dark matter halos.

This work demonstrates that discs, like the dark matter halos they live in, are a fully scaleable population of objects up to the very highest masses.
Our results show that massive spirals are the most efficient galaxies in turning gas into stars in the Universe and that they have very few missing baryons.
We also highlight how the evolutionary paths of discs and spheroids must diverge above a certain mass ($\mstar\simeq10^{10.7} \, \mo$). 
Strong AGN feedback and/or major mergers most likely are responsible for this divergence.
While spheroids grow mostly through episodes of merging and AGN feedback, massive discs grow in near isolation through star formation fed by smooth accretion from their halos, without experiencing disrupting events like major mergers and/or strong AGN feedback.

\section*{Acknowledgements}

The authors thank T.\ Oosterloo, K.\ Spekkens and T.\ van der Hulst for providing reduced \hi\ data of some archival galaxies.
E.M.D.T.\ was supported by the US National Science Foundation under grant 1616177 and by the European Research Council (ERC) under grant agreement no. 101040751.
L.P.\ acknowledges support from the ERC under the European Union Horizon 2020 research and innovation program (grant agreement No. 834148).
M.P.H.\ acknowledges support from NSF/AST-1714828 and grants from the Brinson Foundation.
U.L.\ acknowledges support from from project PID2020-114414GB-100, financed by MCIN/AEI/10.13039/501100011033, from project P20\_00334  financed by the Junta de Andalucia and from FEDER/Junta de Andalucía-Consejer\'ia de Transformaci\'on Econ\'omica, Industria, Conocimiento y Universidades/Proyecto A-FQM-510-UGR20.
Data analysed in this work include multi-band images from WISE, Legacy and Pan-STARSS surveys and spectroscopic datacubes from MaNGA and CALIFA surveys.
This research made use of Photutils, an Astropy package for detection and photometry of astronomical sources.
This work relied on the NASA/IPAC Extragalactic Database, operated by the Jet Propulsion Laboratory, California Institute of Technology, under contract with NASA.

\section*{Data availability}
The data underlying this article will be shared on reasonable request to the corresponding author.


\bibliographystyle{mnras}
\bibliography{biblio_ssvla}

\newpage

\begin{appendix}

\section{Dynamical models for all galaxies}
\label{app:kinmods}

In this Appendix, we show data, kinematical models and rotation curve decompositions for all galaxies analysed in this work. 
For each galaxy, four groups of panels are shown:

\begin{itemize}[itemsep=4pt,leftmargin=0.4cm]
    \item[(a)] Clockwise from the upper-right panel: 1) RGB image in the $g$, $r$ and $z$ bands from the Legacy surveys \citep{Dey+2019} or from Pan-STARRS1 \citep{Chambers+2016}; 2) Surface-brightness profiles derived from $z$-band images; 3) \hi\ column density map; 4) \hi\ velocity field (1st moment). Galaxy centre is denoted as a black cross in the \hi\ total map and velocity field, while the green thick line in the velocity field represents the systemic velocity. The beam of \hi\ data is shown on the column density map in purple. 
    
    \item[(b)] Position-velocity (PV) slices taken along the kinematical major axis (top) and minor axis (bottom).
    Data is shown as blue contours overlaid to the grey colormap, the best-fit kinematical model with \bba\ as red contours. 
    Contour levels are at $2^n \times 2.5\sigma_\mathrm{rms}$, where $\sigma_\mathrm{rms}$ is the rms noise of the data and $n=0,1,2,3,4$. Yellow dots are the best-fit rotation velocities projected along the line of sight (i.e.\ $\vrot\sin i$).
    
    \item[(c)] Best-fit rotation curve (black circles and stars) and rotation curve decomposition. 
    When available, \ha\ rotation curve (stars) is shown together with the \hi\ rotation curve (circles).
    Empty circles denote radii where the rotation velocity is not well determined, which are ignored during the rotation curve decomposition.
    Errors in the rotation velocities include systematic uncertainties on the inclination angle, systemic velocity and galaxy center.
    The contributions of gas, stars and dark matter are shown as blue-dotted, dashed-orange and dashed-dotted purple lines, respectively. 
    The red-solid line denotes the total circular velocity of the model, with the red band indicating the uncertainty.  
    Bottom panel shows the fractional difference between the modelled and the measured circular velocity curve, $\Delta V = (V_\mathrm{mod}-V_\mathrm{obs})/V_\mathrm{obs}$.
    
    \item[(d)] Triangle (corner) plot of the MCMC sampling used for the rotation curve decomposition. Contours plots denote 2D posterior distributions, with contour drawn at 0.5$\sigma$, 1$\sigma$, 2$\sigma$ and 3$\sigma$ confidence levels, while histograms denote the 1D posterior distributions. Full red lines denote the 50th percentile values, dashed black lines are the 15.87th and 84.13th percentiles. The blue band represents the $c_{200}-M_{200}$ relation by \citet{Dutton+2014} used as a prior. We stress that, unlike in \autoref{fig:decomp}, we plot samples of $\log \mstar $ rather than $\Upsilon_\star$.
    
\end{itemize}

\newpage

\begin{figure*}
    \centering
    \includegraphics[width=\textwidth]{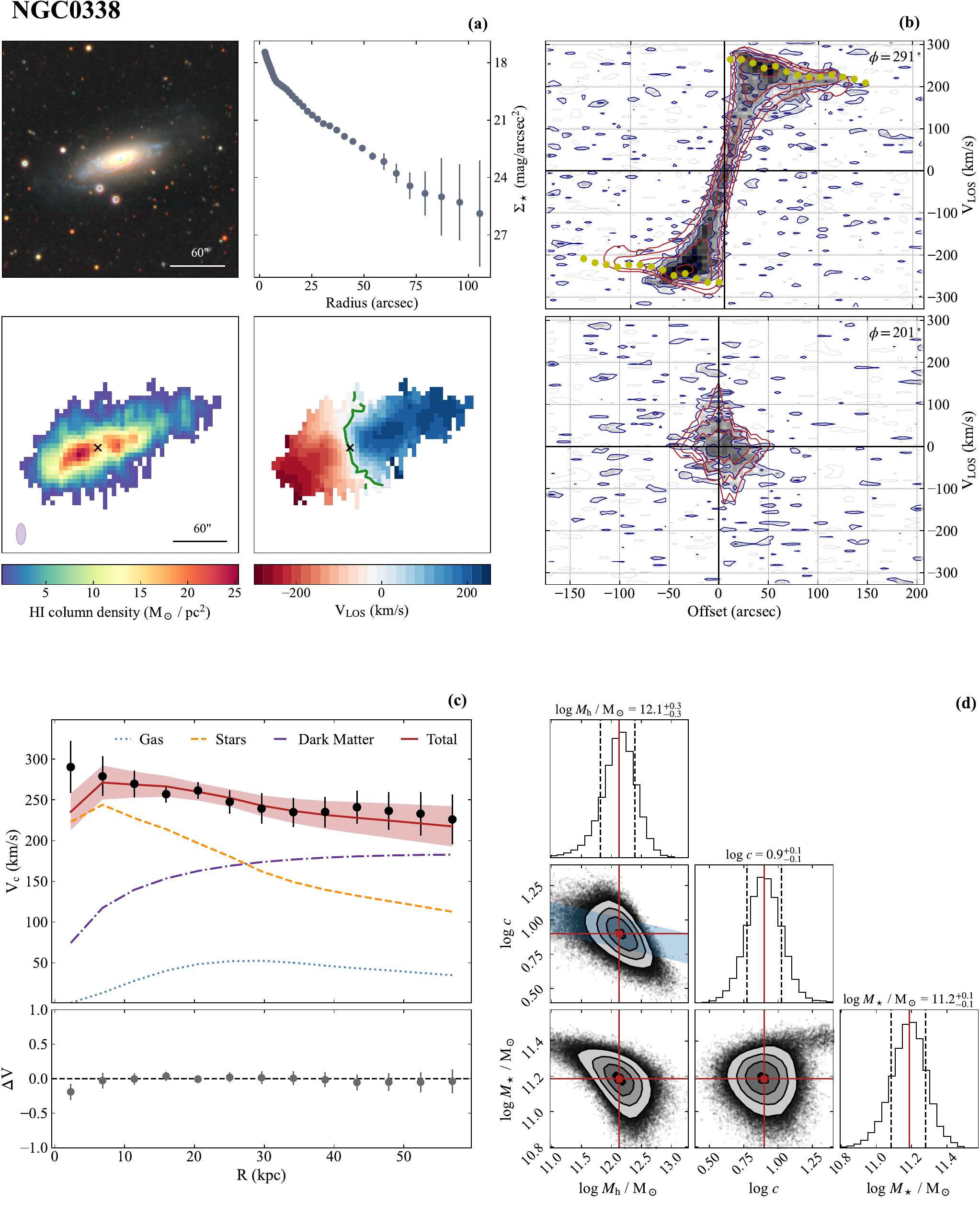}
    \caption{Data and dynamical modelling for galaxy NGC0338. See Appendix~\ref{app:kinmods} for details on single panels.}
    \label{fig:NGC0338}
\end{figure*}
\begin{figure*}
    \centering
	\includegraphics[width=\textwidth]{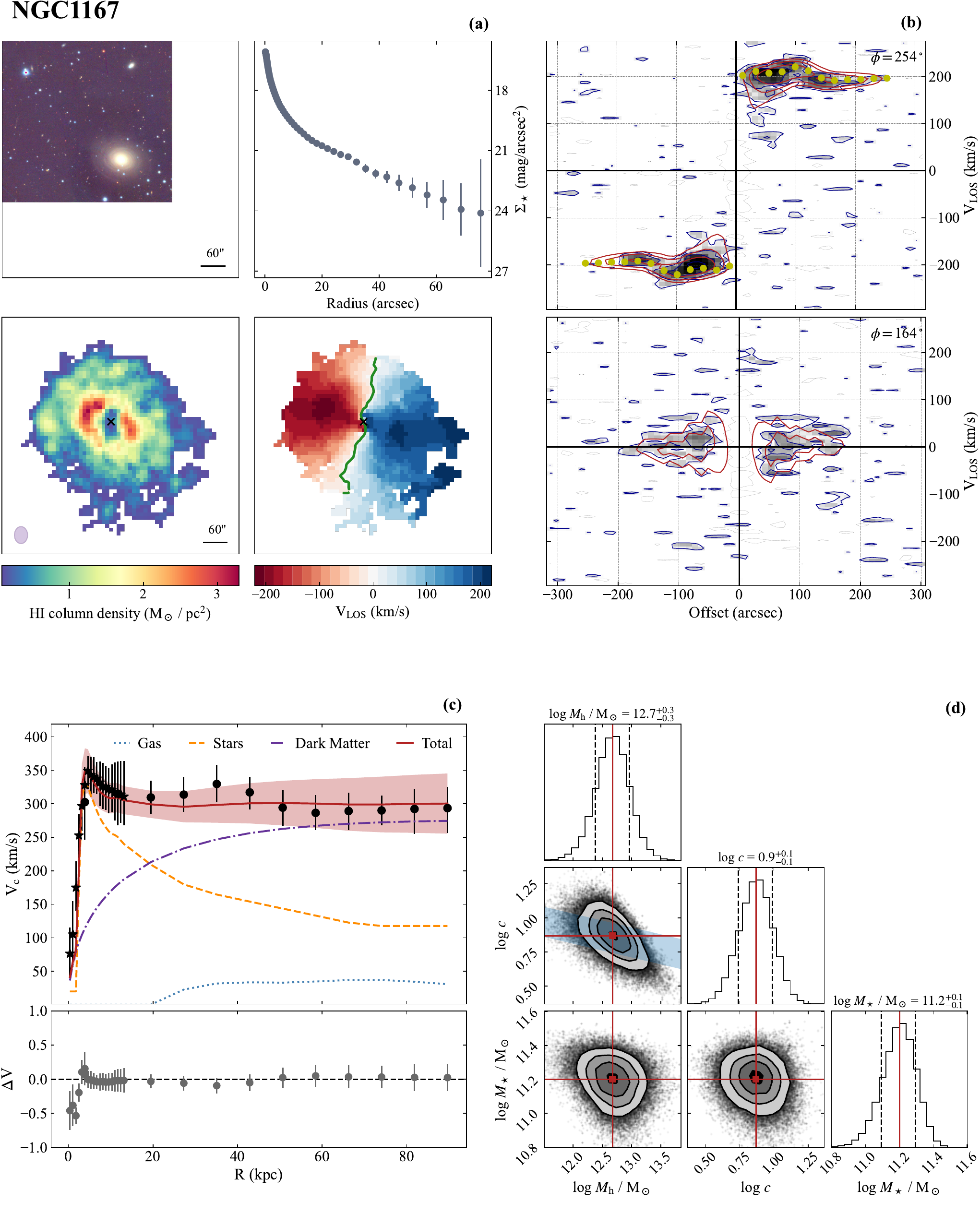}
    \caption{Same as \autoref{fig:NGC0338}, but for galaxy NGC1167.}
\end{figure*}
\begin{figure*}
    \centering
	\includegraphics[width=\textwidth]{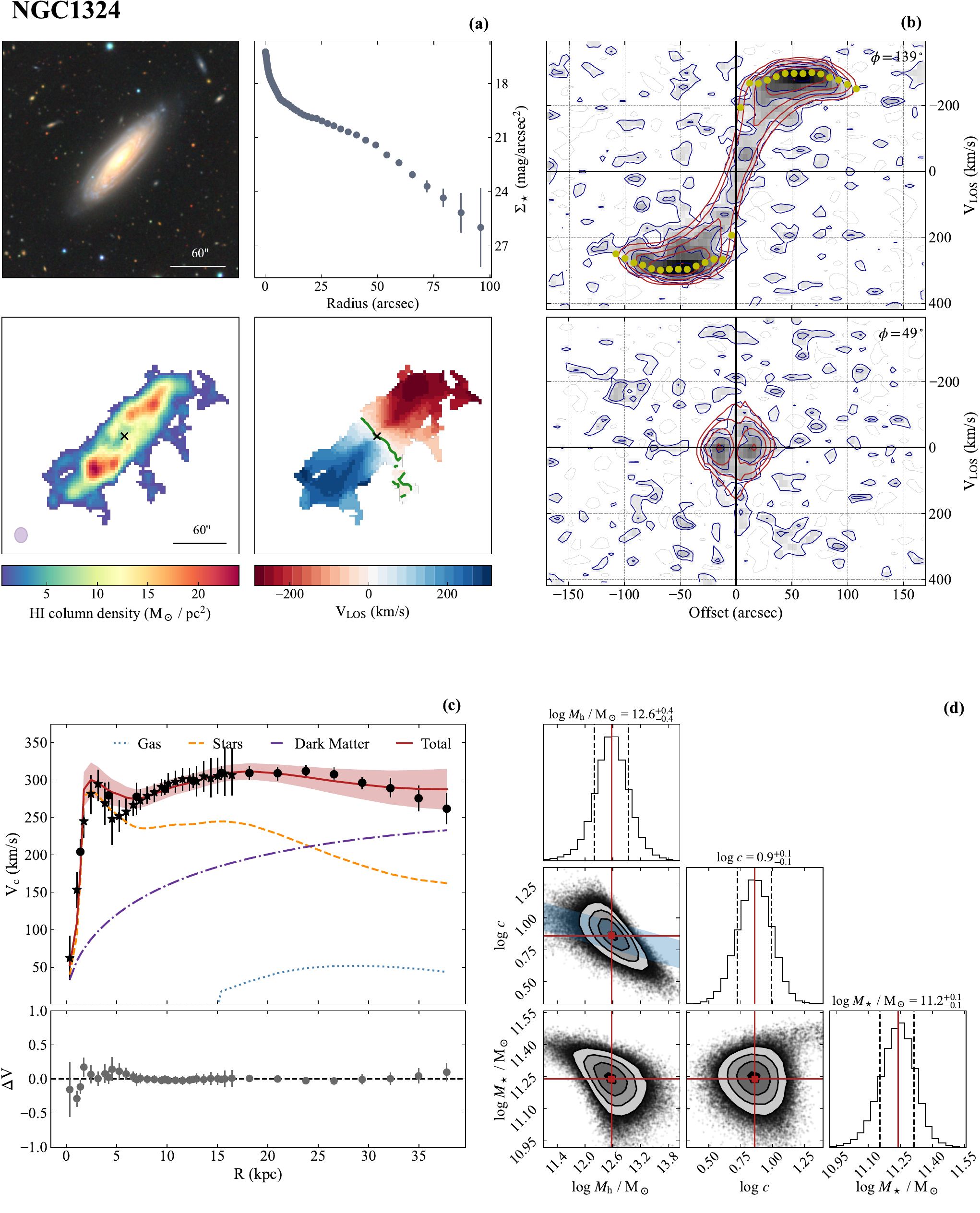}
    \caption{Same as \autoref{fig:NGC0338}, but for galaxy NGC1324.}
\end{figure*}
\begin{figure*}
    \centering
	\includegraphics[width=\textwidth]{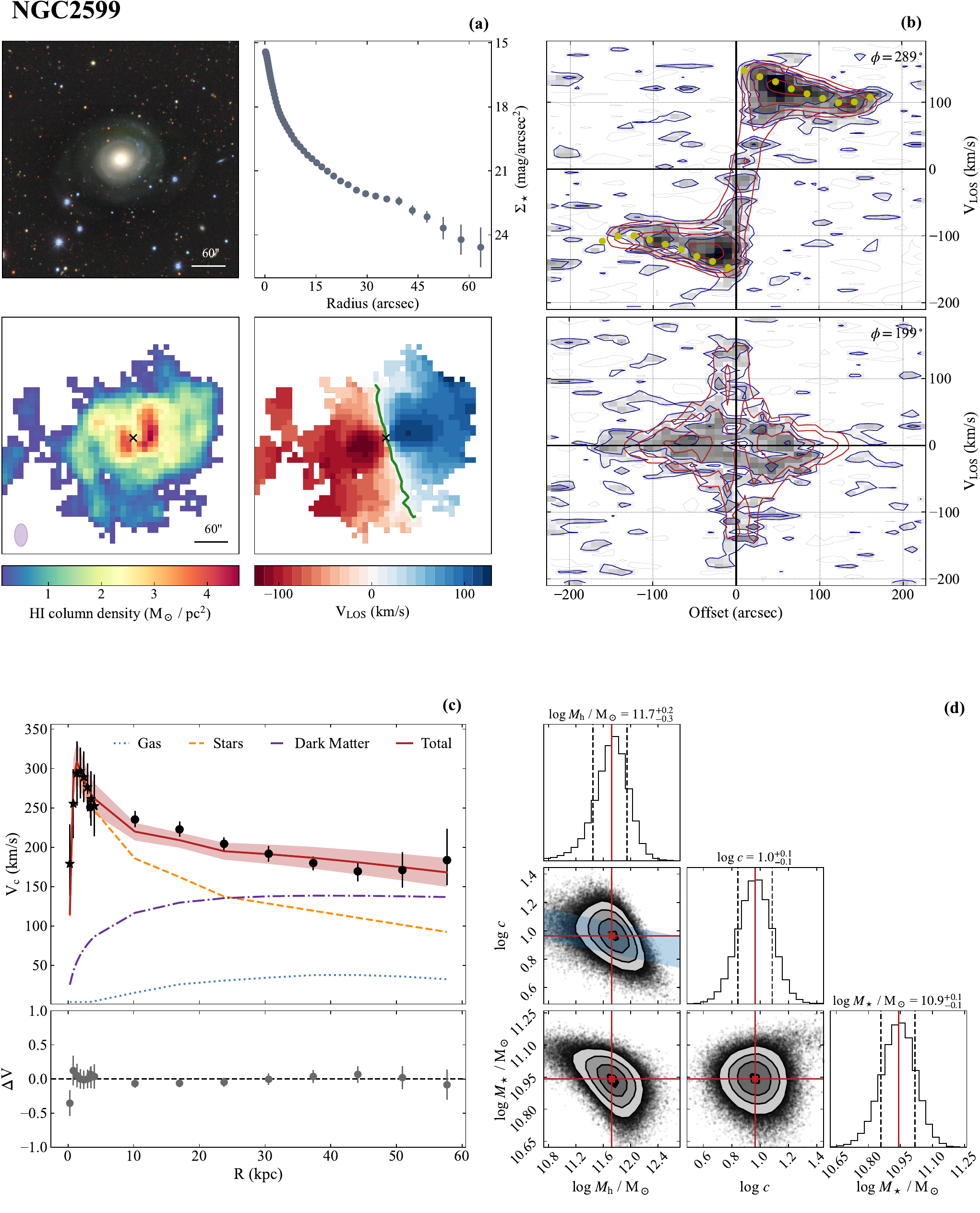}
    \caption{Same as \autoref{fig:NGC0338}, but for galaxy NGC2599.}
\end{figure*}
\begin{figure*}
    \centering
	\includegraphics[width=\textwidth]{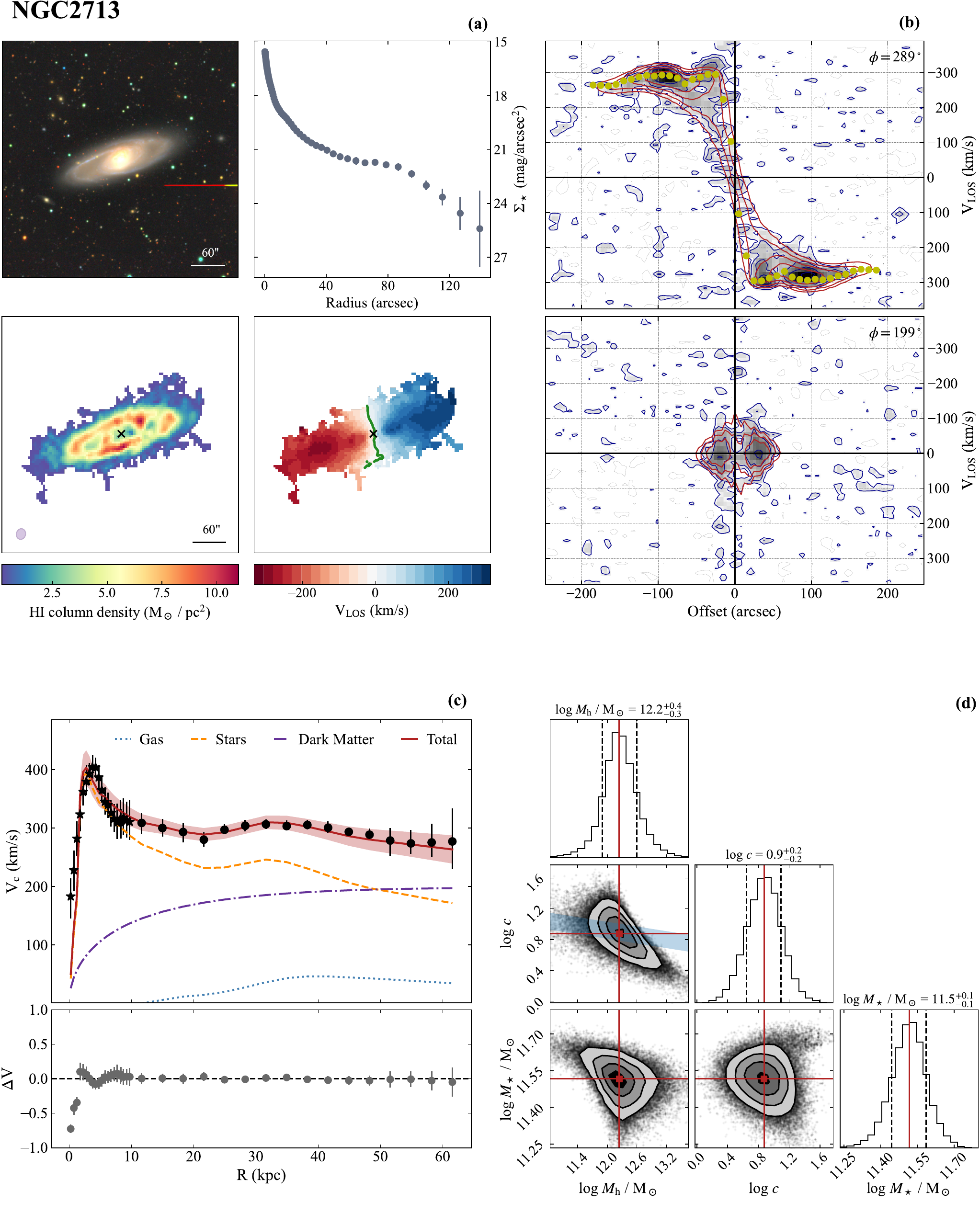}
    \caption{Same as \autoref{fig:NGC0338}, but for galaxy NGC2713.}
\end{figure*}
\begin{figure*}
    \centering
	\includegraphics[width=\textwidth]{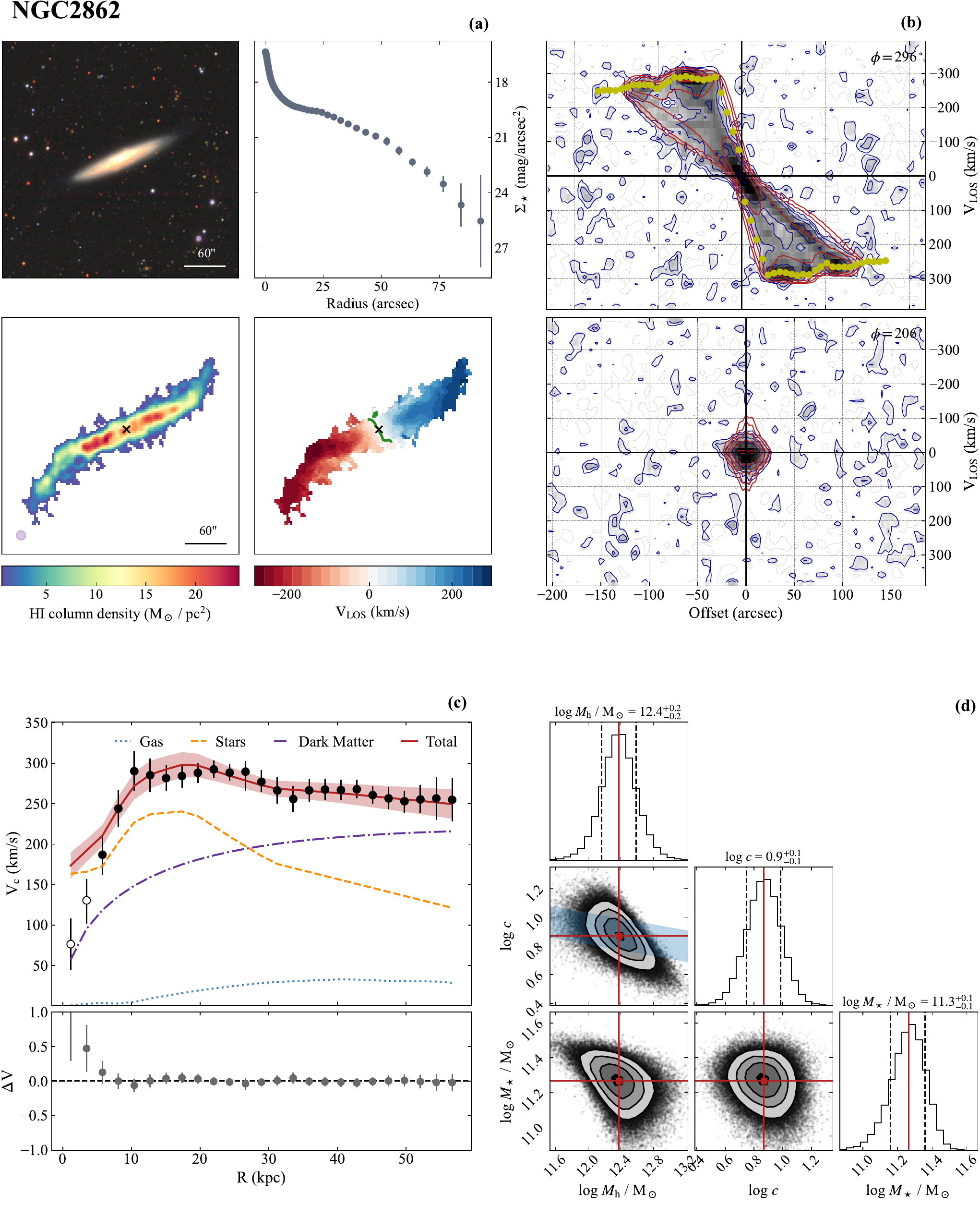}
    \caption{Same as \autoref{fig:NGC0338}, but for galaxy NGC2862.}
\end{figure*}
\begin{figure*}
    \centering
	\includegraphics[width=\textwidth]{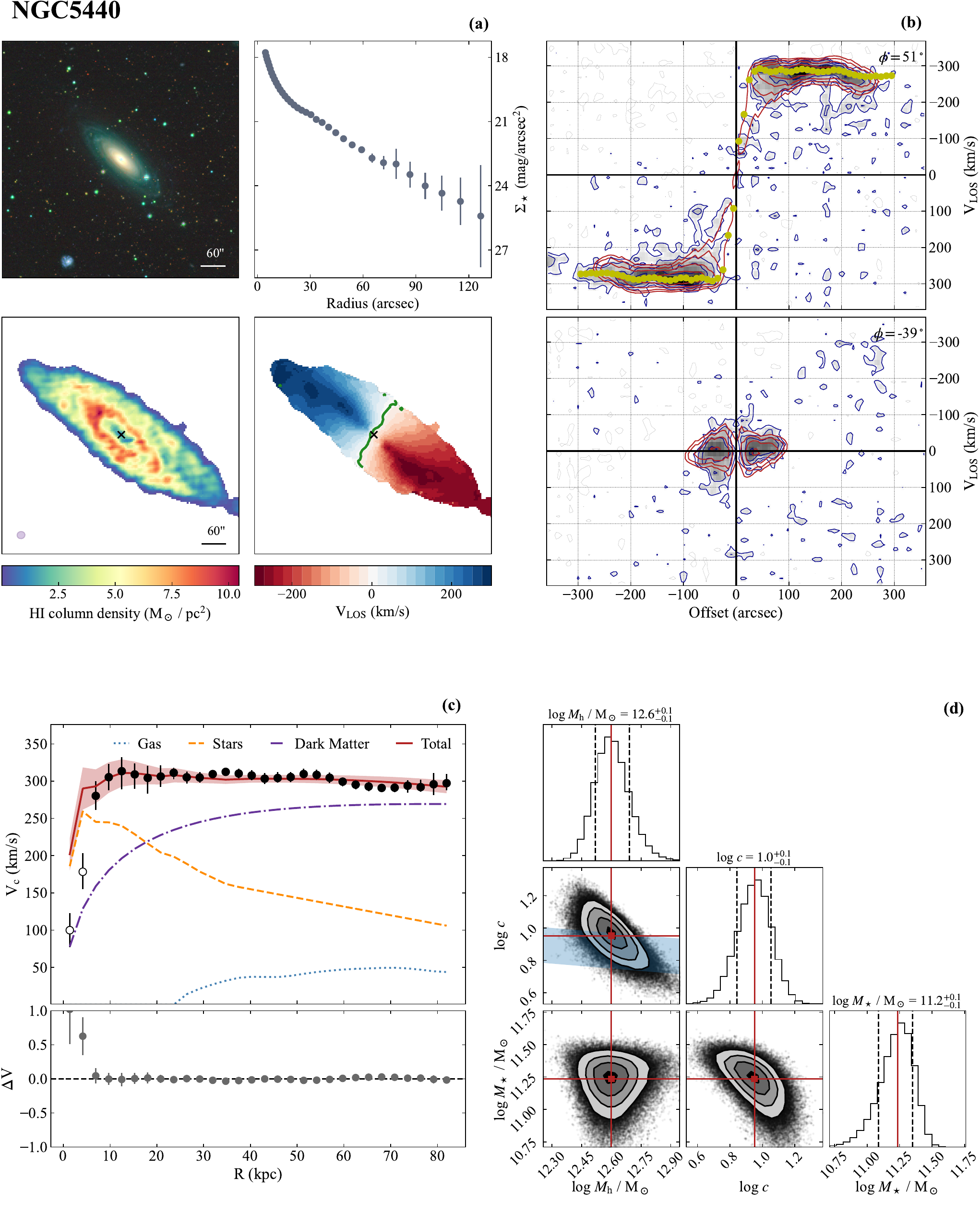}
    \caption{Same as \autoref{fig:NGC0338}, but for galaxy NGC5440.}
\end{figure*}
\begin{figure*}
    \centering
	\includegraphics[width=\textwidth]{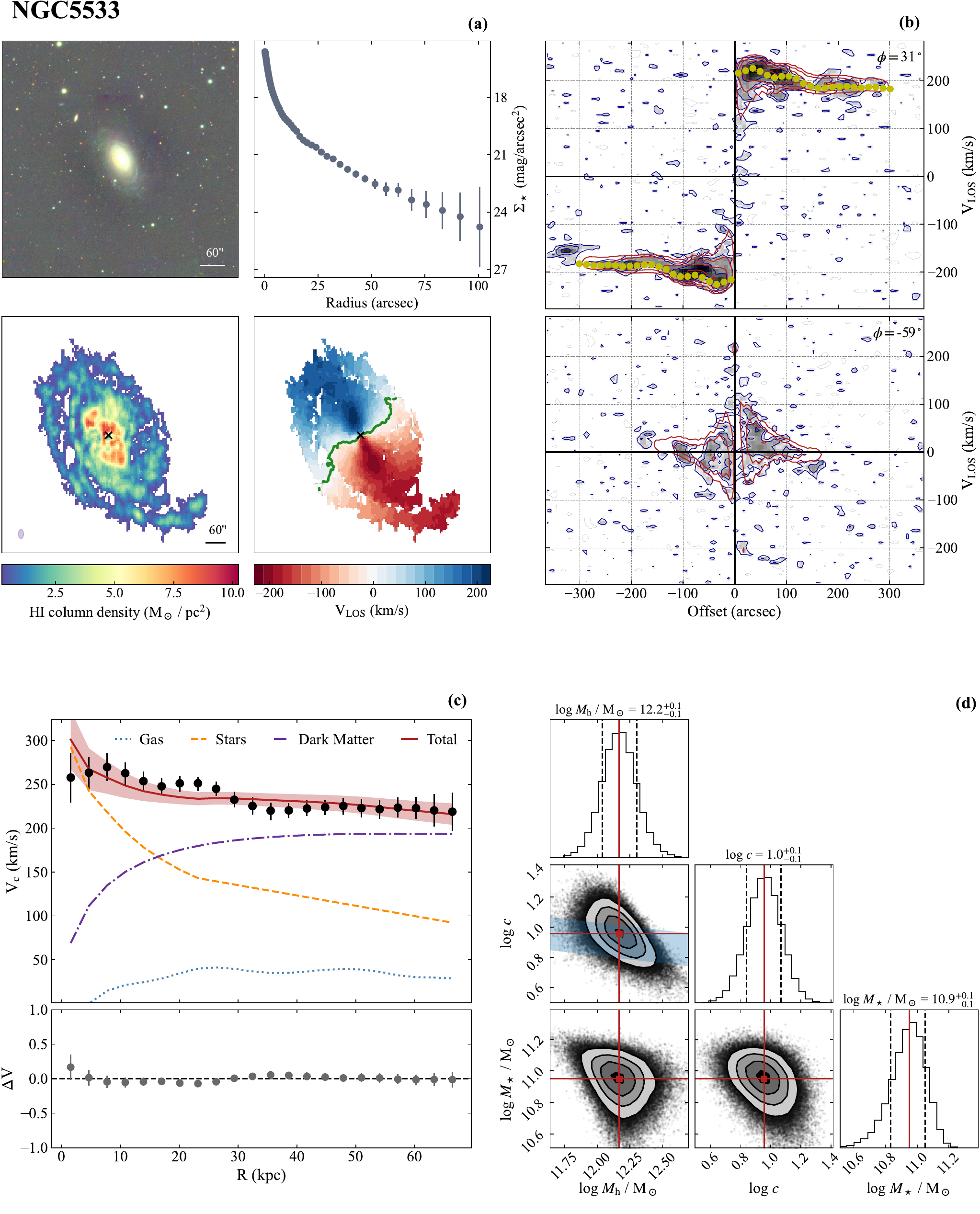}
    \caption{Same as \autoref{fig:NGC0338}, but for galaxy NGC5533.}
\end{figure*}
\begin{figure*}
    \centering
	\includegraphics[width=\textwidth]{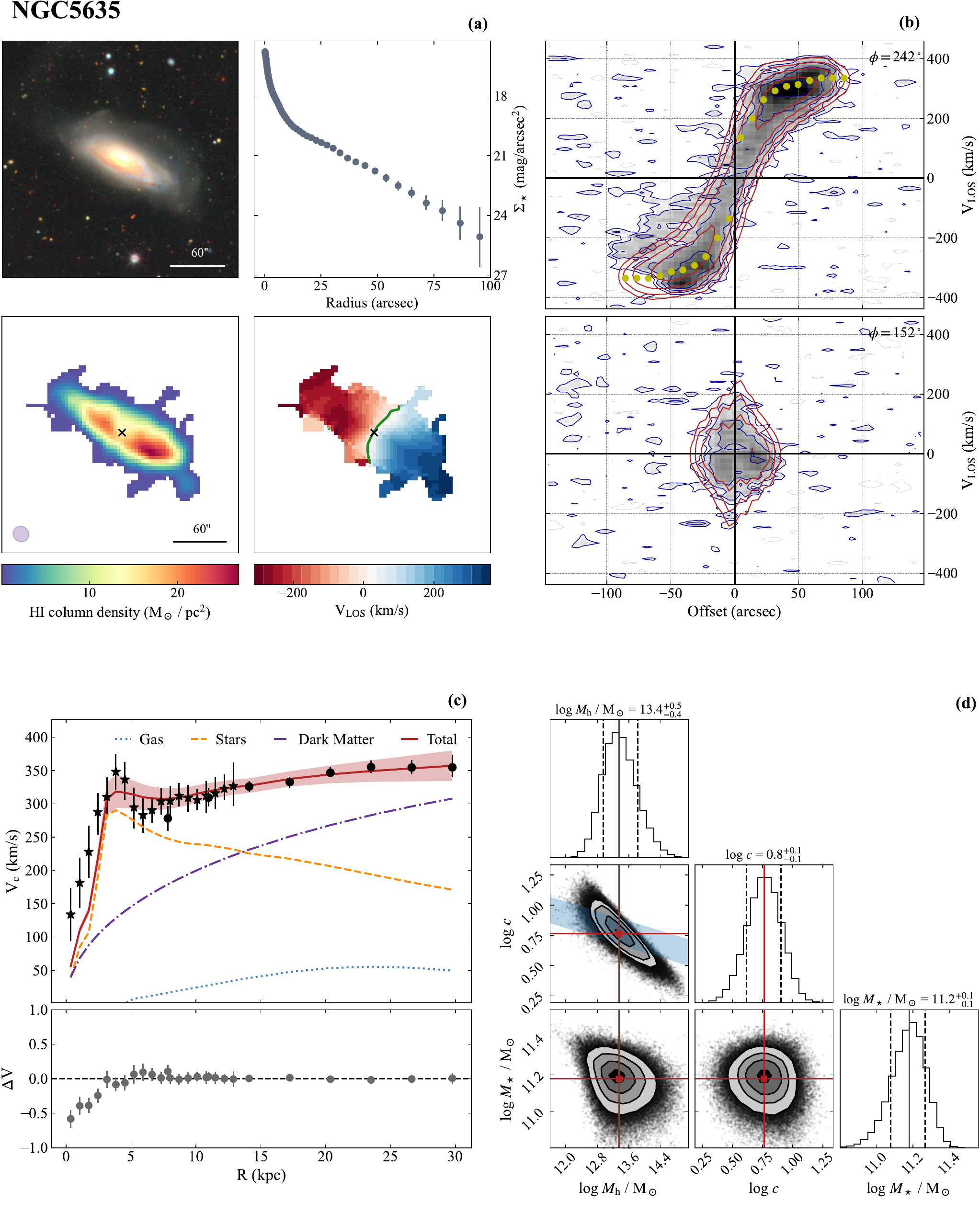}
    \caption{Same as \autoref{fig:NGC0338}, but for galaxy NGC5635.}
\end{figure*}
\begin{figure*}
    \centering
	\includegraphics[width=\textwidth]{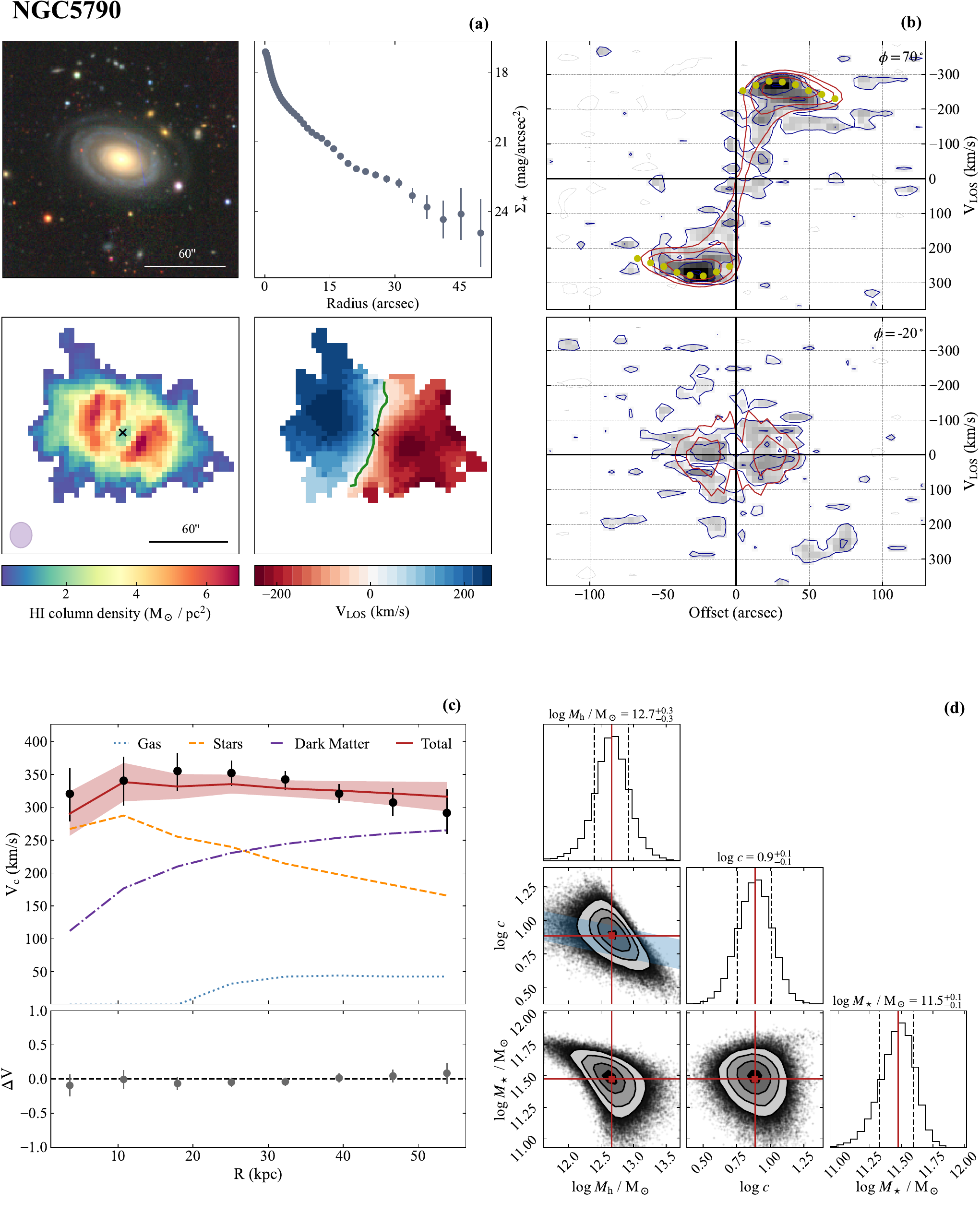}
    \caption{Same as \autoref{fig:NGC0338}, but for galaxy NGC5790.}
\end{figure*}
\begin{figure*}
    \centering
	\includegraphics[width=\textwidth]{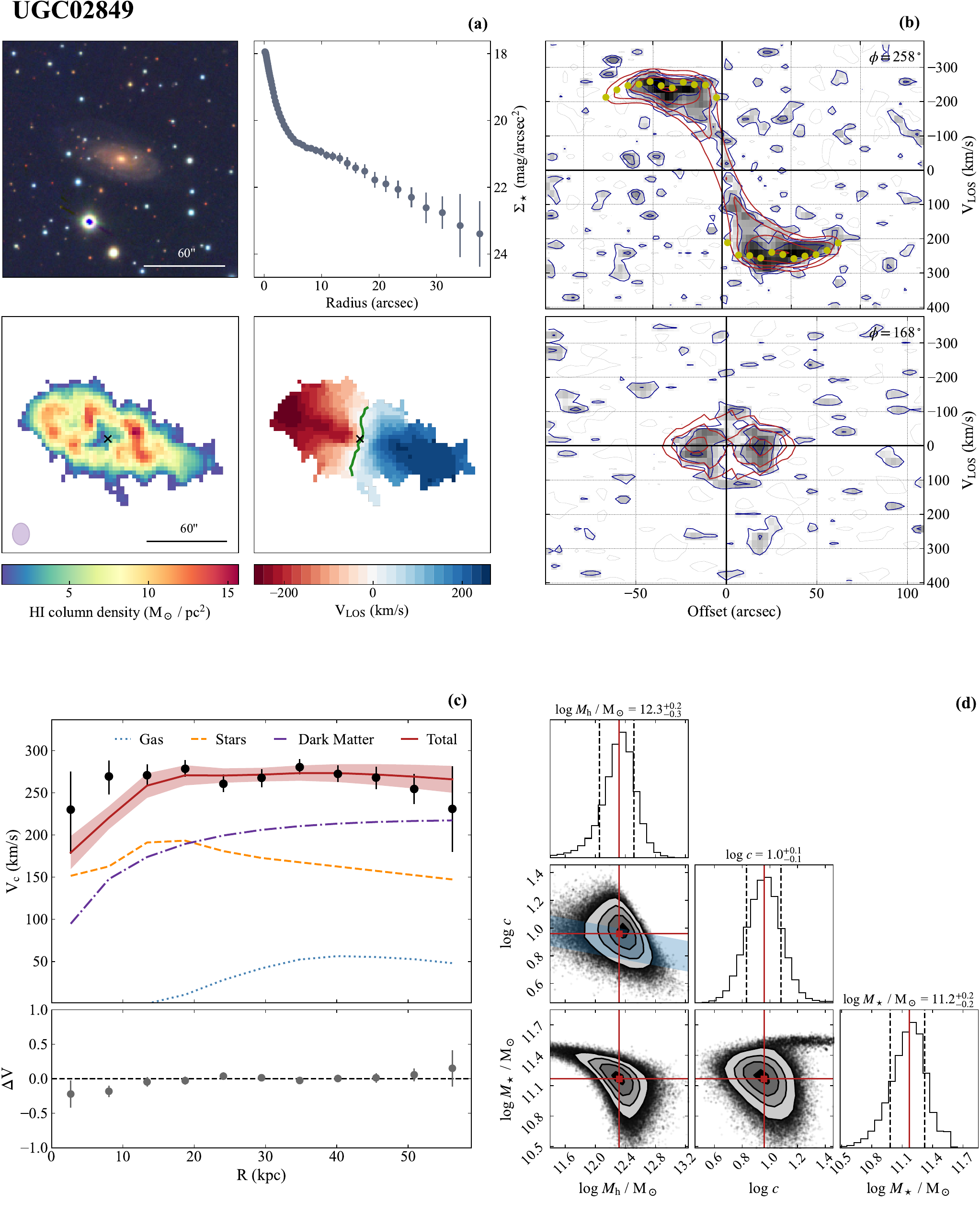}
    \caption{Same as \autoref{fig:NGC0338}, but for galaxy UGC02849.}
\end{figure*}
\begin{figure*}
    \centering
	\includegraphics[width=\textwidth]{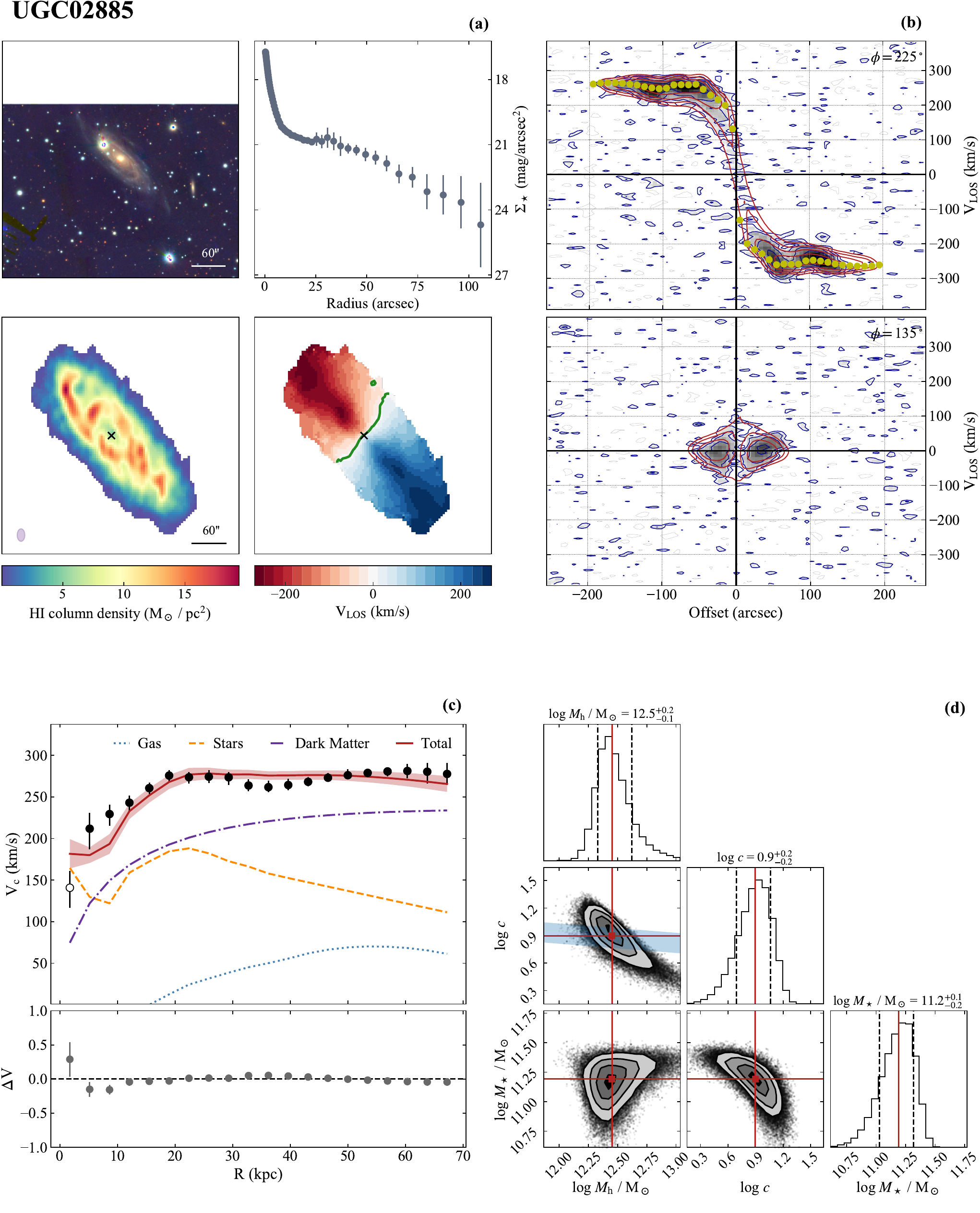}
    \caption{Same as \autoref{fig:NGC0338}, but for galaxy UGC02885.}
\end{figure*}
\begin{figure*}
    \centering
	\includegraphics[width=\textwidth]{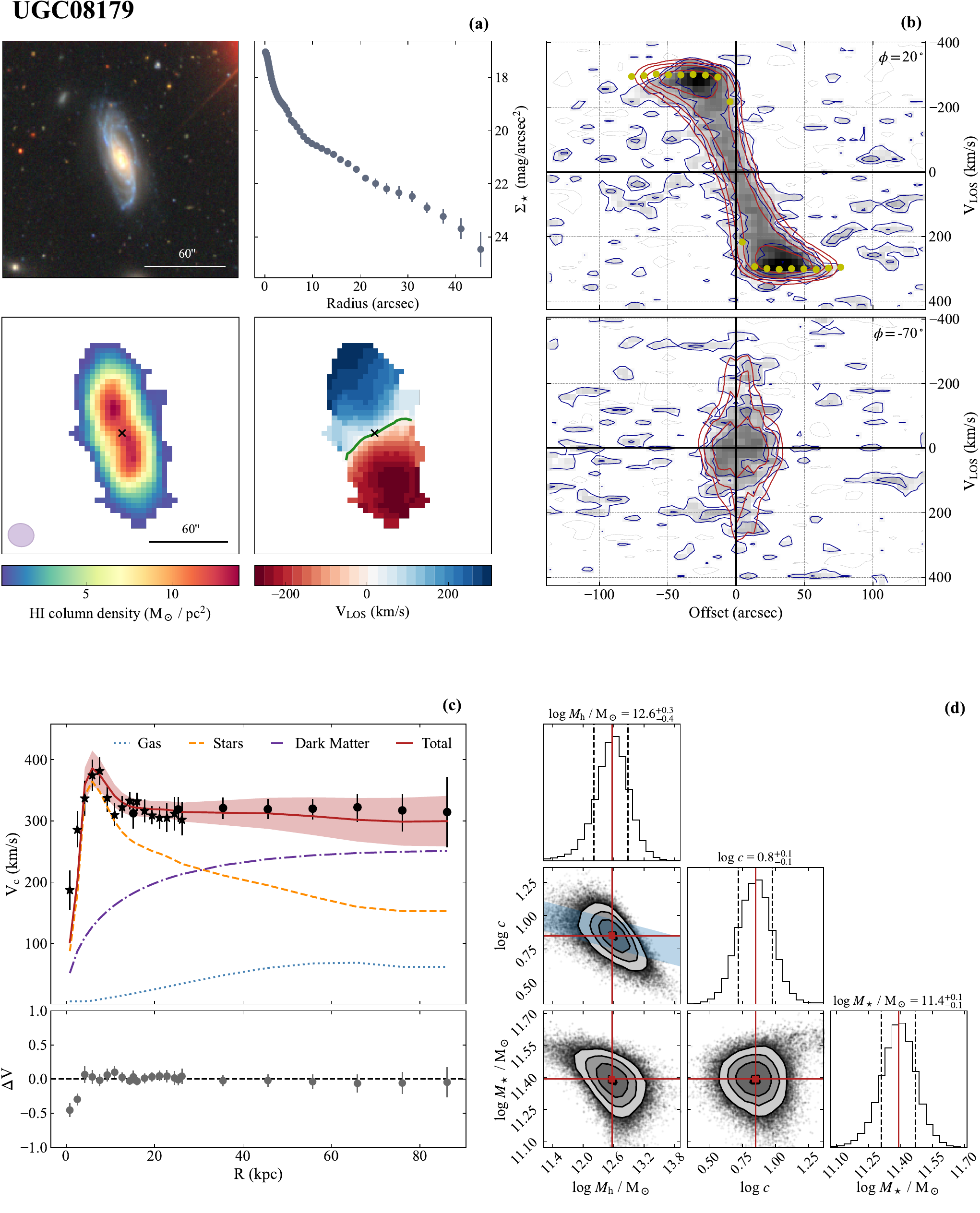}
    \caption{Same as \autoref{fig:NGC0338}, but for galaxy UGC08179.}
\end{figure*}
\begin{figure*}
    \centering
	\includegraphics[width=\textwidth]{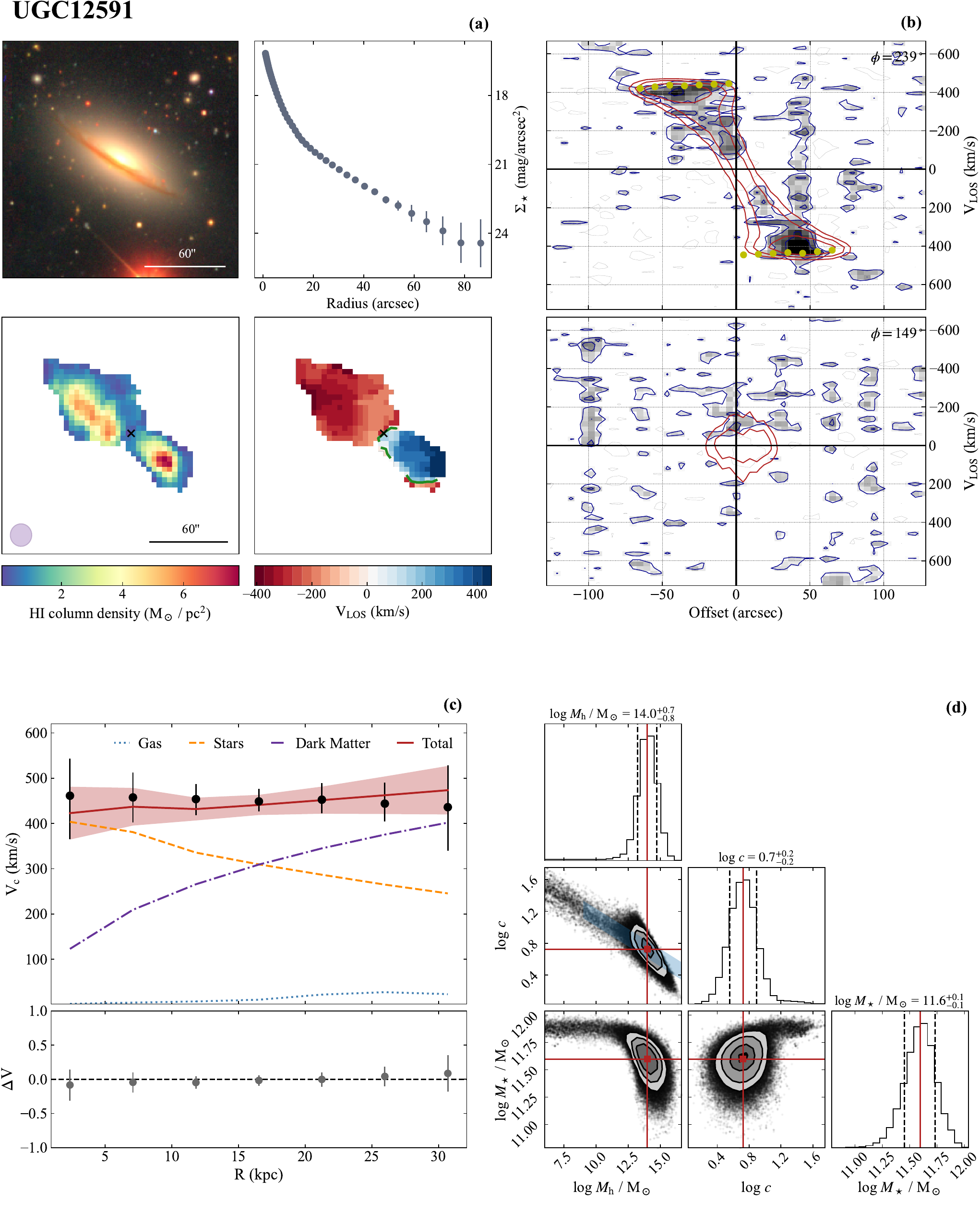}
    \caption{Same as \autoref{fig:NGC0338}, but for galaxy UGC12591.}
\end{figure*}
\begin{figure*}
    \centering
	\includegraphics[width=\textwidth]{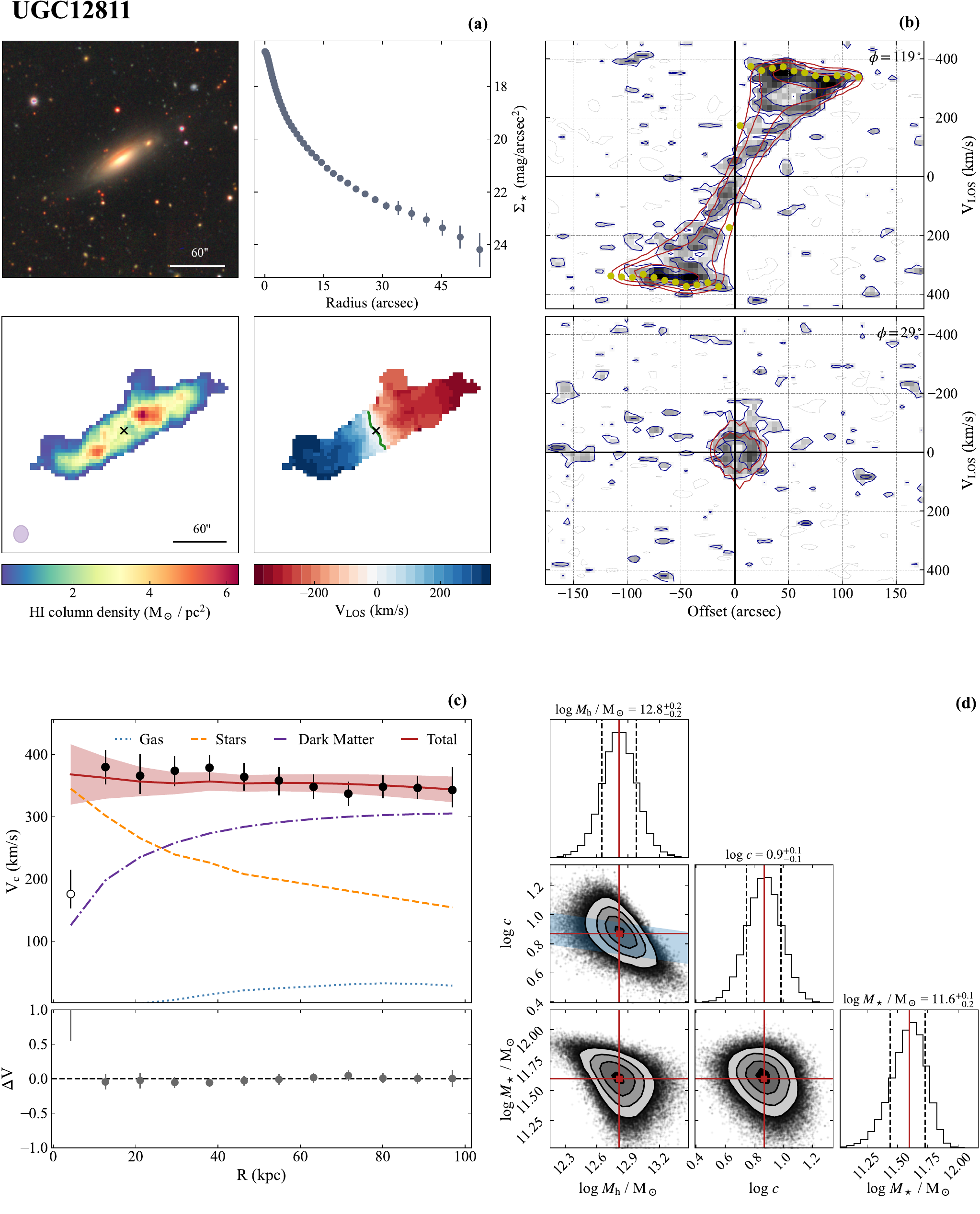}
    \caption{Same as \autoref{fig:NGC0338}, but for galaxy UGC12811.}
\end{figure*}

\section{$W1$-band \mstar vs \mstar\ from rotation curve decomposition}
\label{app:mstar}
\begin{figure}
    \centering
    \includegraphics[width=0.5\textwidth]{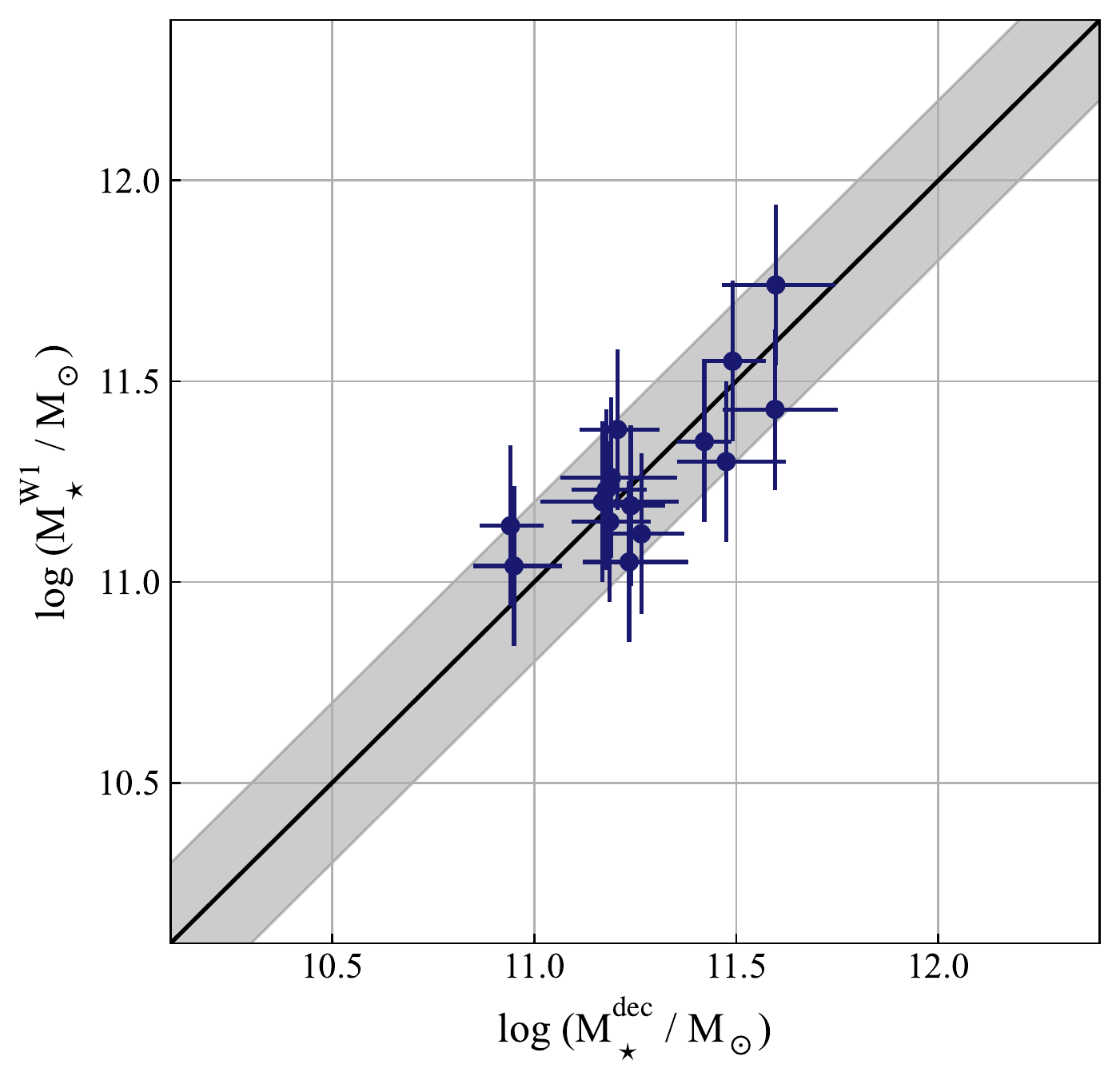}
    \caption{Stellar mass derived from WISE $W1$-band photometry compared to those from rotation curve decomposition for the massive galaxies studied in this work. The black line represents the 1:1 relation, while the grey band denotes the typical 0.2 uncertainty of the photometric stellar masses.}
    \label{fig:stellar_masses}
\end{figure}

In \autoref{fig:stellar_masses}, we compare stellar masses obtained from the total $W1$-band WISE luminosity using a fixed mass-to-light ratio $\Upsilon_{W1}=0.6$ (see Section~\ref{sec:phot}) to those obtained as an output of our rotation curve decomposition procedure (Section~\ref{sec:decomp}).
The grey band denotes the typical 0.2 dex uncertainty of the photometric stellar masses.
We note that, although the two methods are completely independent, the resulting stellar masses are consistent within the errors.

\end{appendix}

\label{lastpage}
\end{document}